\documentclass[lettersize,journal]{IEEEtran}
\usepackage{amsmath}
\usepackage{algorithm}
\usepackage{algorithmicx}
\usepackage{algpseudocode}
\usepackage{enumitem}
\usepackage{subfigure}
\usepackage{graphicx}
\usepackage{booktabs}
\usepackage{multirow}
\usepackage{array} 
\usepackage{tabularx} 
\usepackage{amsfonts}
\usepackage{xcolor}
\usepackage{tcolorbox}
\tcbuselibrary{breakable}
\usepackage{pifont}
\usepackage{url}
\usepackage[table]{xcolor}
\ifCLASSINFOpdf
\else
\fi

\def\BibTeX{{\rm B\kern-.05em{\sc i\kern-.025em b}\kern-.08em
    T\kern-.1667em\lower.7ex\hbox{E}\kern-.125emX}}
\usepackage{balance}

\begin{document}
\title{PDSL: Propagation Dynamics Aware Framework for Source Localization}

\author{Yansong Wang, Qisen Chai, Longlong Lin, Tao Jia

\thanks{This work was supported in part by the Natural Science Foundation of China (No. 72374173), in part by the Natural Science Foundation of Chongqing (No. CSTB2025NSCQ-GPX1082), in part by the Chongqing Graduate Research and Innovation Project (No. CYB23124), in part by the Fundamental Research Funds for the Central Universities (No. SWU-XDJH202303), and in part by the High Performance Computing clusters at Southwest University.}
\thanks{Yansong Wang, Qisen Chai, and Longlong Lin are with the College of Computer and Information Science, Southwest University, Chongqing, 400715, P. R. China. (e-mail: yansong0682@email.swu.edu.cn, cqsllyt@email.swu.edu.cn, longlonglin@swu.edu.cn)}
\thanks{Tao Jia is with the College of Computer and Information Science, Southwest University, Chongqing, 400715, P. R. China, and also with the College of Computer and Information Science, Chongqing Normal University, Chongqing, 401331, P. R. China. (Corresponding author, e-mail: tjia@swu.edu.cn).}}

\markboth{Journal of \LaTeX\ Class Files,~Vol.~14, No.~8, August~2015}%
{Shell \MakeLowercase{\textit{et al.}}: Bare Demo of IEEEtran.cls for IEEE Transactions on Magnetics Journals}

\maketitle


\begin{abstract}
Source localization is a representative inverse inference task in information propagation, aiming to identify the source node or node set that triggers the propagation results based on the observed information. A primary challenge is quantifying the inherent uncertainty between observed outcomes and potential sources. Although deep generative models have partially mitigated this issue, most existing approaches primarily focus on uncertainty induced by network topology, attempting to learn a direct mapping from propagation outcomes to sources based on network structure, while overlooking the additional uncertainty stemming from the highly stochastic nature of the propagation process. To address this limitation, we propose a Propagation Dynamics aware framework for Source Localization (PDSL), a novel method that integrates a deep generative model with propagation dynamics to approximate the source distribution and explicitly mitigate uncertainty arising from diffusion stochasticity. Moreover, we employ Graph Neural Ordinary Differential Equations to model the continuous dynamics of diffusion processes without relying on a predefined diffusion mechanism. Additionally, a matching mechanism is designed to extract relevant data blocks that enhance source generation reliability. Comprehensive experiments on both synthetic and real-world diffusion datasets demonstrate the superior performance of the proposed framework across diverse application scenarios.
\end{abstract}


\IEEEpeerreviewmaketitle

\section{Introduction}\label{sec:intro}
\IEEEPARstart{W}{ith} the exponential growth of online platforms, millions of content pieces are disseminated and reshared by users on a daily basis, rendering information diffusion in social networks a highly intricate and dynamic process \cite{meng2025spreading}. The uncontrolled spread of harmful content, ranging from fake news \cite{lazer2018science, vosoughi2018spread, epstein2023social} to computer viruses \cite{cai2023adam, chen2023modeling, duan2025modeling}, not only misleads the public perception but also inflicts substantial societal and economic consequences. Consequently, mitigating the dissemination of such harmful content has become a key concern for researchers \cite{wei2022modeling,  wang2024modality, liu2025defense}. Within this context, source localization, the reverse problem of dissemination, provides a methodological framework for tracing the origins of propagation by analyzing observed propagation patterns \cite{pinto2012locating}. This paradigm serves as a fundamental measure for reducing the impact of malicious information propagation \cite{dong2019multiple, he2025deciphering}.

Two critical factors, propagation dynamics and network topology, emerge as pivotal elements in this task.
Previous research primarily investigated the relationship between network topology and propagation sources, introducing metrics such as rumor centrality \cite{shah2011rumors}, Jordan centrality \cite{luo2013estimating}, and K-Center \cite{jiang2015k} to quantify the criticality of nodes under various diffusion mechanisms. However, these algorithms enforce relatively rigid preconditions on the information diffusion patterns, which are often challenging to meet in real-world scenarios. To address this issue, some methods \cite{zhu2016information, chang2018maximum, zhang2024multiple} employ the maximum a posterior estimation to identify the source nodes. These probabilistic models are advantageous in capturing and quantifying the dynamics inherent in the information propagation process. Nonetheless, different diffusion mechanisms correspond to distinct parameters, making it challenging to accurately represent all propagation processes with a single probabilistic model. Thus, the generalization of these methods is limited. 
Recently, sensor-based methods \cite{spinelli2017general, wu2020rank, wang2023lightweight} are extensively employed in source localization tasks, typically encompassing two primary stages: sensor deployment and source inference. The first stage designates specific nodes as sensors to record the propagation dynamics of the diffusion process, and the second stage executes inference strategies based on this information, further incorporating the topological properties of the network. The complexity of the model escalates due to the necessity of optimizing both stages concurrently.

\begin{figure}[ht]
    \centering
    \includegraphics[height=0.27\textwidth]{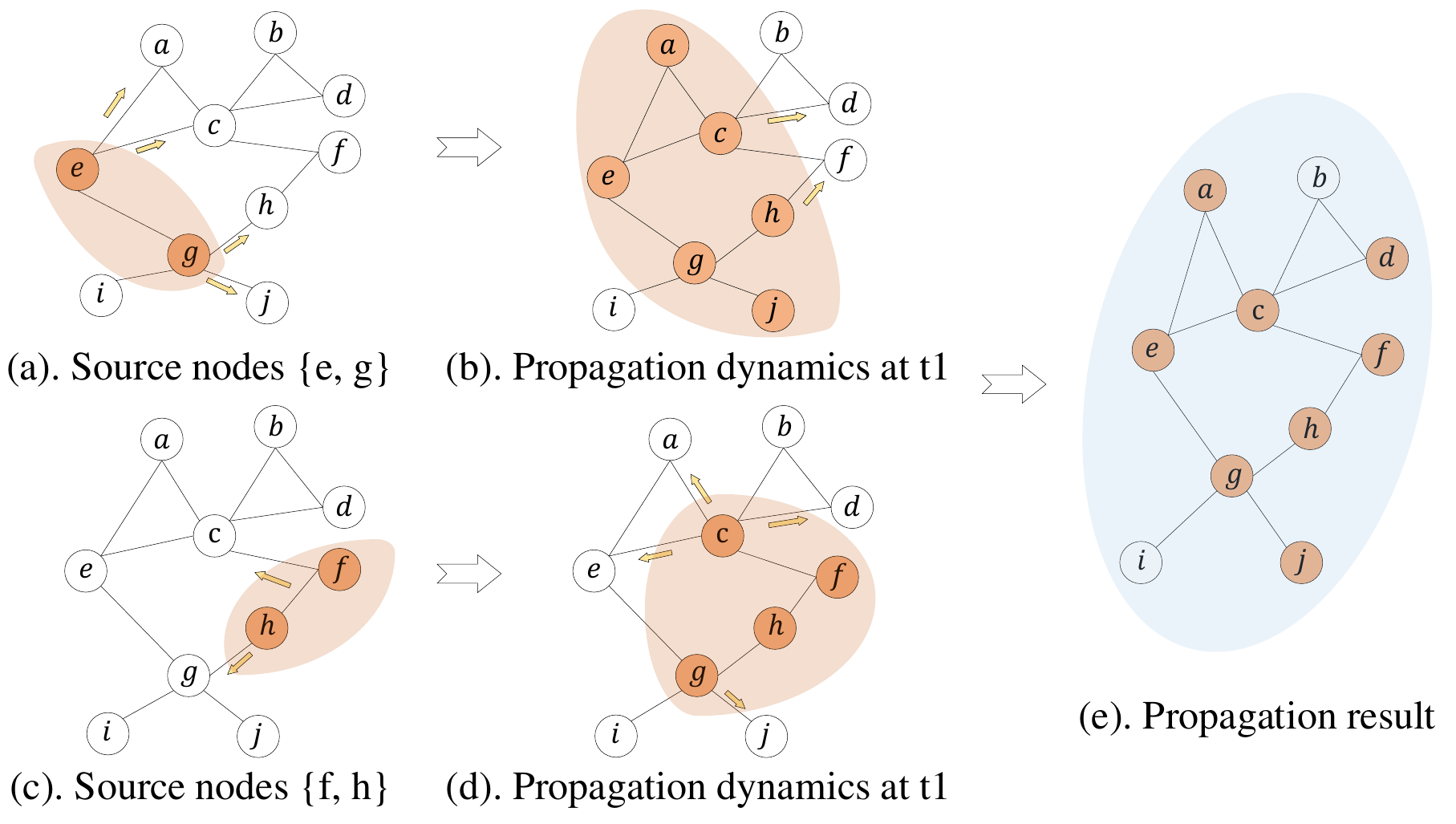}
    \caption{Different source nodes leading to the same propagation result.}
    \label{fig: source-result}
\end{figure}

In the wake of the profound success of deep learning across numerous disciplines \cite{fang2023comprehensive, liu2025coata}, several approaches have begun leveraging neural networks for source localization under diverse diffusion mechanisms \cite{dong2019multiple, shu2021information, yan2024diffusion}. Concurrently, deep generative models, as an important branch of deep learning, offer a robust framework for modeling complex distributions \cite{yang2022learning, saha2025gen}, garnering increasing attention in source localization applications\cite{ling2022source, xu2024pgsl}. 
Despite these advancements, existing methods often focus solely on establishing a direct, static mapping between propagation results and source nodes based on the network topology, failing to adequately address the uncertainty in source localization induced by the stochastic nature of diffusion processes. As delineated in Fig. \ref{fig: source-result}, instances (a) and (c), different sources denoted by the sets $\{e, g\}$ and $\{f, h\}$ respectively, result in an identical propagation result (e). If the propagation dynamics (b) and (d) are neglected during modeling, the method can only generate a single source distribution and does not effectively account for all possible source node combinations. Consequently, there is a need for methods that accurately identify the source nodes by considering the stochastic nature of diffusion processes.
In this pursuit, we are mainly faced with the following challenges.
1) The origins of stochasticity are multidimensional, including network topology, propagation dynamics, and node attributes. The effective integration of these elements into the framework is a major challenge. 
2) The state transition of nodes is governed by a complex interplay of node interactions, temporal dependencies, and external environmental factors, typically exhibiting continuous evolutionary patterns rather than discrete state jumps. Developing methods to precisely model these continuous transition dynamics is another challenge.

In response to the aforementioned challenges, we propose a Propagation Dynamics aware framework for Source Localization (PDSL), a novel method that synergistically integrates propagation dynamics and network topology with deep generative models to approximate source distributions, and incorporates Graph Neural Ordinary Differential Equations (Graph Neural ODEs) to continuously model node state evolution throughout the forward diffusion process. This architecture allows the model to learn the intrinsic continuous evolution laws of information diffusion. Specifically, PDSL consists of two steps: the source quantification and the forward propagation. In the first step, we employ conditional deep generative models, constrained by limited observed propagation dynamics, to infer the initial source probability distribution. It allows us to model the probability distribution of these latent sources rather than making a deterministic point estimate. Subsequently, the forward propagation implements continuous diffusion modeling through Graph Neural ODEs, generating simulated propagation results that serve as feedback to iteratively refine the source distribution estimation. Their integration enables the model to jointly optimize inverse inference and forward simulation, which is critical for reducing uncertainty in source localization.
The principal contributions of this work are delineated as follows:
\begin{itemize}[leftmargin=1em]
    \item The framework integrates propagation dynamics into deep generative models for source localization. This hybrid approach enables probabilistic inference of source distributions while avoiding critical limitations that either overlook propagation dynamics or rely on oversimplified probabilistic assumptions.
    \item The introduction of Graph Neural ODEs provides a continuous paradigm for modeling node state evolution during information diffusion without requiring predefined diffusion mechanisms. This allows for a more flexible simulation of information dissemination within the network.
    \item The inference phase incorporates a result similarity based matching strategy to initialize the source localization, which leverages prior knowledge from the training set to provide initialization closer to the true solution space, accelerating optimization convergence.
    \item We construct synthetic diffusion datasets on four real-world networks and one synthetic network using different diffusion mechanisms, and conduct experiments on these datasets. In addition, we perform further validation on two real-world diffusion datasets involving changes in network topology. The results demonstrate improved performance of PDSL over state-of-the-art baselines.
\end{itemize}

In summary, the rest of the paper is organized as follows. Section \ref{sec:related} reviews the related works. The details and sufficient analyses of this framework are shown in Section \ref{sec:method}. In Section \ref{sec:experiment}, extensive simulations and corresponding experiments are conducted. Finally, we conclude the model and discuss future directions of this topic in Section \ref{sec:conclusion}.

\section{Related Work}\label{sec:related}

\begin{table*}[htbp]
\caption{Overview of Typical Source Localization Methods.}
\centering
\renewcommand{\arraystretch}{1.2} 
\footnotesize
\begin{tabularx}{\textwidth}{>{\centering\arraybackslash}m{1.4cm}>{\centering\arraybackslash}m{0.4cm}X>{\raggedright\arraybackslash}m{3.8cm}}
\hline
\textbf{Type} & \textbf{Ref} & \multicolumn{1}{c}{\textbf{Contributions and Differences}} & \multicolumn{1}{c}{\textbf{Limitations}} \\ \hline

\multirow{8}{=}{\centering Topology Based} 
& \cite{shah2011rumors} & Pioneering work. Uses the SI model to construct a source estimator based on a novel topological measure called rumor centrality. 
& \multirow{8}{=}{1. Rely on predefined, rigid diffusion mechanisms. \\ 2. Struggle to capture the stochastic and continuous nature of real-world propagation. \\ 3. Prone to creating fake paths that deviate from actual dynamics.} \\ \cline{2-3}
& \cite{luo2014identify} & Employs the Jordan Center to investigate source detection under the SI model. & \\ \cline{2-3}
& \cite{li2023rumor} & Introduces infection potential energy to address path inaccuracy, emphasizing the directional influence of infected nodes. & \\ \cline{2-3}
& \cite{ma2024dislpsi} & Focuses on signed networks; integrates structural balance theory to capture dynamics across positive and negative edges. & \\ \cline{2-3}
& \cite{jiang2024source} & Proposes a signed dynamic message passing algorithm with positive/negative transmission rates to improve accuracy. & \\ \hline

\multirow{8}{=}{\centering Probabilistic Model Based} 
& \cite{zhu2016information} & Derives the Maximum A Posteriori (MAP) estimator of the source for tree networks under the Independent Cascade (IC) model. 
& \multirow{8}{=}{1. Often assume simple or specific propagation models, restricting generalizability. \\ 2. Fail to effectively quantify source uncertainty, leading to sub-optimal solutions in ill-posed settings.} \\ \cline{2-3}
& \cite{chang2018maximum} & Detects a single source in weighted graphs under the SI model based on likelihood approximation and MAP estimation. & \\ \cline{2-3}
& \cite{dawkins2021diffusion} & Introduces a principled statistical framework to construct a confidence set for the source node. & \\ \cline{2-3}
& \cite{gong2024hmsl} & Incorporates higher-order Markov properties via a reaction-diffusion process to solve the path length underestimation problem. & \\ \cline{2-3}
& \cite{zhou2025multi} & Combines graph representation learning with Bayesian optimization to estimate parameters from a single snapshot. & \\ \hline

\multirow{8}{=}{\centering Diffusion Model Based} 
& \cite{ling2022source} & Combines forward diffusion estimation with deep generative models to approximate the source distribution. 
& \multirow{8}{=}{1. Often fail to incorporate underlying propagation dynamics as a continuous process. \\2. A gap exists between discrete denoising and actual continuous state evolution of nodes.} \\ \cline{2-3}
& \cite{yan2024diffusion} & Constructs a discrete denoising diffusion model using reversible residual network blocks based on MPNN relationships. & \\ \cline{2-3}
& \cite{huang2023two} & Uses temporal cascade information and coarse-grained initialization to tackle scalability and mapping ambiguities. & \\ \cline{2-3}
& \cite{hou2025generalized} & Redefines forward diffusion to learn unbiased noise in a self-supervised manner, removing reliance on explicit source labels. & \\ \cline{2-3}
& \cite{chen2025structure} & Addresses data scarcity by using structure-prior biased initialization and propagation-enhanced conditional denoisers. & \\ \hline

\multirow{7}{=}{\centering Sensor Deployment Based} 
& \cite{spinelli2017general} & Provides a framework for source location via both static and dynamic sensor placement during and after epidemics. 
& \multirow{7}{=}{1. Lack of interconnectivity between the deployment and inference stages. \\2. High maintenance costs and heavy dependence on accurate timestamps. \\3. Limited practical application in evolving environments.} \\ \cline{2-3}
& \cite{cheng2022path} & Deploys observers in multiplex networks to record directions and handle path uncertainty. & \\ \cline{2-3}
& \cite{hou2024random} & Proposes a random full-order neighbor selection strategy for rapid deployment in large-scale networks. & \\ \cline{2-3}
& \cite{zhao2025enhanced} & Partitions the network into communities to ensure timely responses in source localization. & \\ \cline{2-3}
& \cite{zhu2022locating} & Identifies source nodes with local maximum labels through an iterative process using snapshots and recorded paths. & \\ \cline{2-3}
\hline

\multirow{8}{=}{\centering Neural Network Based} 
& \cite{dong2019multiple} & First GCN-based model; transforms integer labels into vectors and incorporates multi-order neighbor information. 
& \multirow{8}{=}{1. Focus mostly on static mappings. \\2. Struggle to model continuous evolutionary patterns of diffusion. \\3. Suffer from high uncertainty.} \\ \cline{2-3}
& \cite{wang2022invertible} & Develops an invertible graph diffusion model with error compensation and validity-aware layers. & \\ \cline{2-3}
& \cite{hou2024new} & Integrates user profiles to develop a user-centric framework based on real-world cascades. & \\ \cline{2-3}
& \cite{bao2024graph} & Uses graph contrastive learning with propagation-stochasticity-aware data augmentation to reduce label reliance. & \\ \cline{2-3}
& \cite{sun2025trace} & Geometric perspective; establishes a structural Schr\"{o}dinger bridge on Riemannian manifolds via geodesic matching.  & \\ \hline

\end{tabularx}
\label{tab:related work}
\end{table*}

Source detection is a complex inverse problem, which aims to use key information such as observed network topology and node states to track back and identify the initial source that triggered the propagation process. It involves many applications, such as epidemic prevention \cite{shah2020finding, ru2023inferring} and opinion guidance \cite{li2012multiple, zang2015topic}. Existing source localization methods can be broadly categorized into five paradigms: topology based methods, probabilistic model based methods, diffusion model based methods, sensor deployment based methods, and neural network based methods. Table \ref{tab:related work} provides a structured overview of the most representative works across different categories.

\textbf{Topology Based Methods.} This category of techniques primarily identifies sources by mining structural network characteristics and integrating them with deterministic or heuristic propagation models to reconstruct diffusion paths. \cite{shah2011rumors} models rumor spreading in a network using the SI model and constructs an estimator for the rumor source based on a novel topological measure called rumor centrality. This work is regarded as pioneering in the task of source localization in social networks. Similarly, \cite{luo2014identify} employs the Jordan Center to investigate source detection using the same path under the SI model. \cite{jiang2015k} proposes a K-center method to identify multiple diffusion sources and their corresponding infection regions in altered networks, along with the SI model to predict spreading time and a heuristic algorithm for estimating the number of sources. Recently, \cite{li2023rumor} introduces a novel source localization method based on infection potential energy to address the inaccuracy of diffusion paths in traditional methods. Unlike purely statistical methods, this approach emphasizes the directional influence of infected nodes and employs network pruning to optimize the source search process. In light of the prevalent adversarial and friendly relationships in social media, research on signed networks explores how structural balance theory and edge sign features impact information tracing. \cite{ma2024dislpsi} introduces a specialized framework for source localization in signed social networks, which integrates the theory of structural balance to capture the nuanced dynamics of information flow across positive and negative edges. \cite{ma2024source} introduces an observer selection optimization based on effective distance and uses a reverse propagation algorithm to locate sources, finding that a higher proportion of positive edges facilitates accurate localization. Similarly, \cite{jiang2024source} proposes a signed dynamic message passing algorithm. This method modifies the SIR model to incorporate positive/negative transmission rates and utilizes edge attributes to significantly improve localization accuracy in structurally balanced networks. Despite their simplicity, these methods often rely on predefined, rigid diffusion mechanisms. They struggle to capture the stochastic and continuous nature of real-world propagation, leading to fake paths that deviate from actual transmission dynamics.

\textbf{Probabilistic Model Based Methods.} Probabilistic inference methods treat source localization as a posterior maximization problem, utilizing statistical frameworks to identify the most likely origin within highly stochastic diffusion processes. \cite{zhu2016information} studies information source detection under the Independent Cascade model \cite{goldenberg2001talk}, and derives the maximum a posteriori estimator of the source for tree networks. \cite{chang2018maximum} studies the problem of detecting a single information source in a weighted graph from the perspective of likelihood approximation and derives a method for detecting a single information source under SI model based on a maximum a posteriori estimator. \cite{chai2021information} proposes a novel efficient algorithm to estimate the information source and diffusion time simultaneously. This method represents time-varying networks with a time-aggregated graph and uses the SIR model for node diffusion dynamics. \cite{dawkins2021diffusion} introduces a statistical inference framework that constructs the confidence set for the source node in a more natural and principled way. \cite{zhang2024multiple} adopts Bayesian optimization to promote efficiency and reveal a relationship between the node set and the observation. \cite{gong2024hmsl} challenges the conventional first-order Markov assumption. This method formulates a reaction-diffusion process to capture the higher-order Markov properties of propagation, effectively solving the problem where first-order networks underestimate path lengths and activation times. Addressing the multi-source localization challenge, \cite{zhou2025multi} combines graph representation learning to capture latent topological features with Bayesian optimization to efficiently estimate diffusion parameters and time from a single snapshot. Many probabilistic approaches assume simple or specific propagation models, which restricts their generalizability across different scenarios. Furthermore, they often fail to effectively quantify source uncertainty, leading to sub-optimal solutions in ill-posed settings. 

\textbf{Diffusion Model Based Methods.} Diffusion models simulate a reverse denoising process to reconstruct initial states, effectively quantifying uncertainty and solving ill-posed inverse problems. \cite{ling2022source} proposes a probabilistic model that combines forward diffusion estimation with deep generative models to approximate the diffusion source distribution. \cite{yan2024diffusion} constructs a discrete denoising diffusion model that introduces a reversible residual network block based on the relationship between diffusion phenomena and message-passing neural networks. However, direct application of diffusion models faces challenges like computational cost and ill-posedness. To overcome this, \cite{huang2023two} proposes a two-stage optimization framework. It utilizes temporal cascade information to resolve mapping ambiguities and employs a coarse-grained initialization to tackle the scalability issues typically associated with iterative diffusion sampling on graphs. \cite{hou2025generalized} introduces a generalized framework that redefines the forward diffusion process to learn unbiased noise in a self-supervised manner. This allows for tracing propagation back to the source without relying on explicit source labels, enhancing generalizability across different models. Addressing data scarcity, \cite{chen2025structure} utilizes a structure-prior biased diffusion initialization and a propagation-enhanced conditional denoiser. This enables effective knowledge transfer from synthetic data to real-world scenarios with limited propagation data. Despite their power, existing diffusion frameworks often fail to incorporate the underlying propagation dynamics as a continuous process, leading to a gap between discrete denoising and the actual continuous state evolution of nodes.

\textbf{Sensor Deployment Based Methods.} Sensor based techniques focus on strategically selecting observation nodes to capture temporal or directional metadata, enabling precise localization while minimizing data acquisition costs. \cite{spinelli2017general} proposes a comprehensive framework for source location that includes both static and dynamic sensor placement. This framework enables source location both during the active spread and after the epidemic has spread the entire network. Addressing the cost of maintaining sensors, \cite{spinelli2017back} pioneered the online approach for source localization, addressing the high maintenance costs of static sensors. The framework iteratively chooses the optimal location for new sensors based on real-time infection states. \cite{paluch2020optimizing} introduces a novel deployment strategy called collective betweenness, which optimizes traditional betweenness centrality to provide the highest quality of source location, particularly in scenarios where spreading is highly stochastic and unpredictable. For more complex topologies like multiplex networks, \cite{cheng2022path} applies source centrality theory and deploys observers to record spreading directions, effectively handling the high uncertainty of spreading paths in multi-layer structures. To address the limitations of traditional sensor-based methods that rely solely on single-aspect knowledge, \cite{zhao2024mase} transforms the source localization problem into a multi-attribute decision-making problem, proposing a general framework that integrates multi-attribute source estimators and source dimension estimators, thereby enhancing the method's flexibility and scalability. Timeliness is also a critical factor. \cite{hou2024random} proposes a random full-order neighbor selection strategy that enables rapid sensor deployment across large-scale networks, while \cite{zhao2025enhanced} adopts a community-based perspective to partition the network into multiple clusters, thereby ensuring timely responses in source localization. Similarly, \cite{shi2025community} proposes a framework that integrates community partitioning with adaptive observer deployment. It utilizes contrastive learning for partitioning and employs an early source estimation strategy to enhance both efficiency and accuracy. Some works concentrate on the localization task under low infection rate scenarios \cite{wang2022rapid, zhu2022locating}. \cite{wang2022rapid} proposes a greedy full-order neighbor location method. This approach considers the greedy strategy during sensor deployment and considers the relationship between observed time and the actual propagation path in the source inference process, allowing it to pinpoint the source within a small area early on. \cite{zhu2022locating} uses the obtained snapshots and recorded paths, an iterative process is conducted until the source nodes with local maximum labels are identified. For improved accuracy in large-scale networks with less computational complexity, \cite{wang2023lightweight} proposes a greedy-coverage-based rapid source localization method. This method greedily deploys sensors to rapidly achieve wide area coverage at a low cost and executes source inference in a small, early-infected area. \cite{liu2023diffusion} develops a percolation-based evolutionary framework for optimizing the sensor set, with the aim of minimizing the candidate sources. Traditional sensor methods are often static and lack interconnectivity between the deployment and inference stages. Moreover, the high cost of maintaining sensors and the heavy dependence on accurate timestamps limit their practical application in evolving network environments.

\textbf{Neural Network Based Methods.} Neural networks leverage powerful feature extraction capabilities to automatically learn complex non-linear mappings between observed snapshots and source nodes, overcoming the dependence on specific analytical models. \cite{dong2019multiple} firstly introduces a Graph Convolutional Networks based Source Identification model. This model constructs node representation by transforming the integer label into the vector and incorporating multi-order neighbor information. \cite{shu2021information} identifies information sources using a multi-channel graph neural network framework. The node channel leverages the network structure to represent each node as an embedding vector, while the edge channel converts the network into a line graph to extract edge features. These features are then aggregated to estimate the probability of each node being the source. \cite{wang2022invertible} introduces a graph residual scenario to make graph diffusion models invertible, develops an error compensation mechanism, and creates validity-aware layers to ensure inferred sources are feasible. \cite{xu2024pgsl} leverages graph neural networks to encapsulate propagation patterns between observed diffusion and sources of high uncertainty, addressing the inherent uncertainty in inverse graph diffusion and generating expressive posteriors with smooth and invertible transformations. To tackle the data limitation problem, \cite{wang2024joint} extends the scope to cross-platform scenarios. The framework mines implicit knowledge from cascades with similar topics across different platforms. It features a dual-channel structure with a self-loop attention GCN and a KL regularization module to constrain the latent distribution of source probabilities. Building on snapshot observations, \cite{hou2025fgssi} proposes an inductive localization framework based on a sequence-to-sequence architecture that leverages snapshot observations to address the influence of user interactions in time-varying contagion scenarios, while ensuring transferability across different contexts. Conversely, \cite{hou2024new} builds on real-world propagation cascades, integrating user profiles to develop a user-centric framework for source localization. To alleviate the heavy reliance on labeled data in supervised GNN-based methods, \cite{bao2024graph} explores self-supervised learning for source localization. This work proposes a graph contrastive learning framework that employs a propagation-stochasticity-aware data augmentation strategy and a feature enrichment module, achieving superior performance and transferability across different networks. From a geometric perspective, \cite{sun2025trace} proposes a transferable model that establishes a structural Schrödinger bridge on the Riemannian manifold. By using geodesic bridge matching, it avoids expensive iterative procedures and effectively models the map between source and final distributions. Unlike methods relying on timestamps, \cite{ma2026rumor} develops a direction-based pruning framework. By leveraging the transmission direction toward observed nodes, the framework significantly narrows the candidate source range without requiring prior knowledge of propagation dynamics. While GNNs capture topological features well, most existing methods focus on static mappings. They struggle to model the continuous evolutionary patterns of information diffusion and often suffer from high uncertainty.

\section{Method}\label{sec:method}
\begin{figure*}
    \centering
    \includegraphics[width=0.95\textwidth]{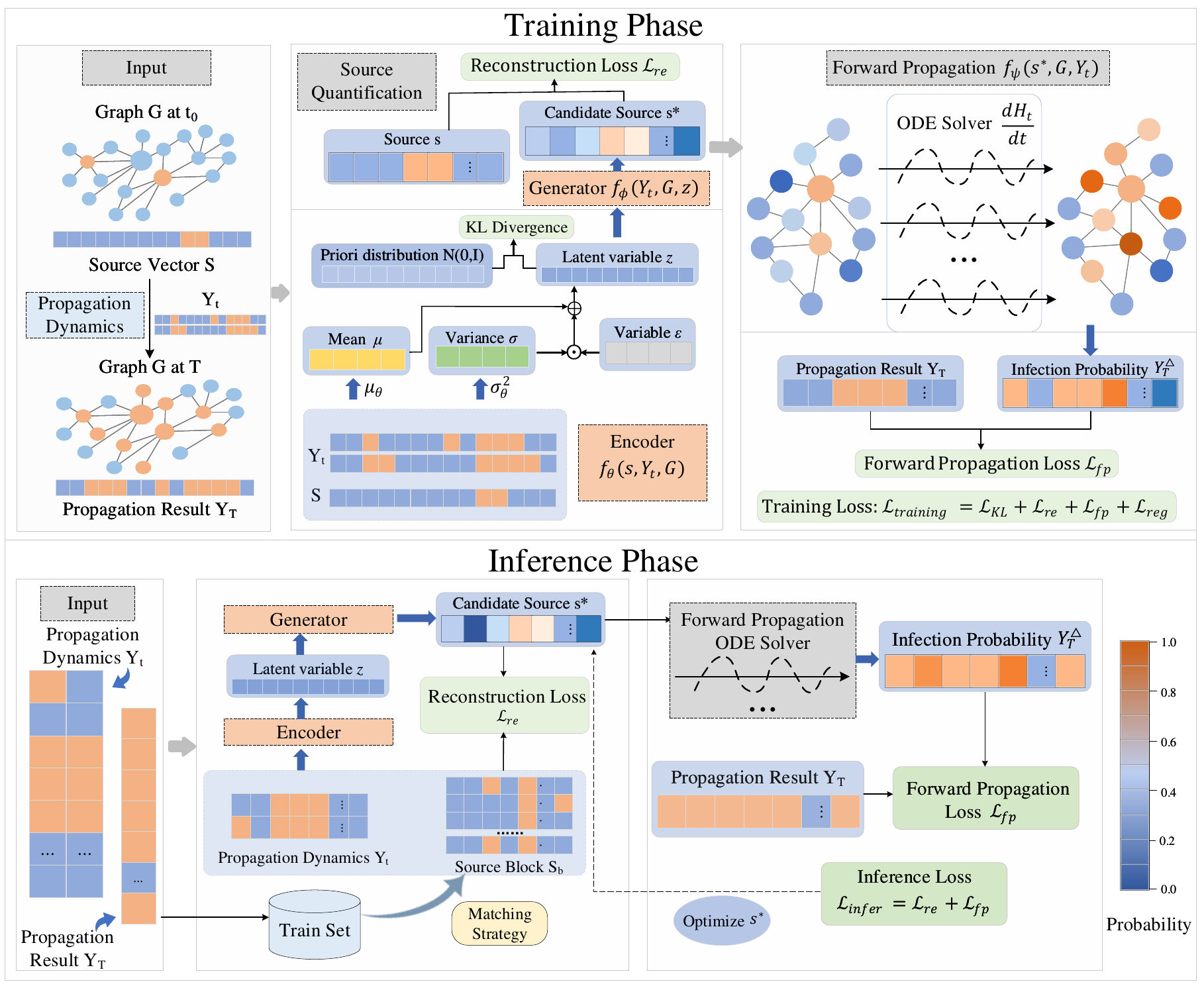}
    \caption{The architectural overview of our framework. During the training phase, the proposed framework takes graph topology $G$, source vector $s$, snapshot vectors $Y_t$, and propagation result $Y_T$ as input. The source quantification encompasses two key components: (1) the Encoder that translates the inputs, excluding $Y_T$, into a continuous latent variable $z$. (2) the Generator that produces a potential source vector $s^*$, conforming to a specific distribution based on $z$. The forward propagation employs a Graph Neural ODEs to derive the propagation output $Y_T^\triangle$. The loss $\mathcal{L}_{training}$ is calculated to optimize the model parameters. In the inference phase, the framework ingests graph topology $G$, snapshot vectors $Y_t$, and propagation result $Y_T$. It initiates by aligning the known propagation result $Y_T$ with the training dataset to identify the most analogous data block $S_b$. Subsequently, the framework leverages the trained source quantification and forward propagation modules to generate the propagation output $Y_T^\triangle$. Ultimately, the gradient values derived from the loss function are utilized to iteratively refine the potential source vector $s^*$.}
    \label{fig:framework}
\end{figure*}

This section first presents the problem formulation, followed by a detailed exposition of the modules. The overall architecture of PDSL is illustrated in Fig. \ref{fig:framework}.
\subsection{Problem Formulation}
Given a graph $G = (V,E)$, where $V = \{v_1, ..., v_n\}$ is the set of all nodes and $E$ is the set of all edges. Suppose one message propagates in the network $G$ according to a specific diffusion mechanism $D$, and each node has two states: infected state and susceptible state. We model the state of the nodes at any given time $t$ using a snapshot vector $Y_t \in \{0,1\}^n$ where $Y_{t,i} = 1$ means node $v_i$ is infected in time $t$, it is susceptible otherwise. $s \in \{0,1\}^n$ is a source vector, $s_i = 1$ if $v_i$ is infected and $s_i = 0$ otherwise. The diffusion process starts at timestamp $0$ and ends at timestamp $T$. The propagation result is $Y_T \in \{0,1\}^n$. Traditional forward prediction focuses on the prediction of $Y_T$ through the source vector $s$ and propagation dynamics $\{Y_t\}_{t=1}^{T-1}$ (for simplicity, we use $Y_t$ to represent propagation dynamics unless otherwise specified). While the source localization problem is defined as reconstructing a source vector $\hat{s} \in \{0,1\}^n$ through the propagation result $Y_T$, limited observed information $Y_t$ and diffusion mechanism $D$,
    $\hat{s} = {\arg\max}_{s} p(s | Y_T, Y_t, G, D)$,
where $p$ is a conditional probability of $s$ given $Y_T$, $Y_t$, $G$ and $D$.

While sustained research efforts to develop diffusion mechanisms \cite{cencetti2023distinguishing} that adhere to real-world diffusion patterns, the inherent stochasticity and multifactorial nature of information spread \cite{hodas2014simple} make it difficult to define with a single, definitive model. Therefore, we focus on source localization without relying on a specific diffusion mechanism,
\begin{equation}
    \hat{s} = \mathop{\arg\max}\limits_{s} p(s | Y_T, Y_t, G).
\label{equ1}
\end{equation}
This implies our method takes a data-driven perspective where we assume nothing about the underlying diffusion mechanism $D$ other than it being driven by the topology $G$, and we use limited past known observations $Y_t$ and the propagation result $Y_T$ to train our model.

\subsection{Source Quantification}
Intuitively, propagation results $Y_T$ depend on the source $s$, network topology $G$ and propagation dynamics $Y_t$, we extend Equation (\ref{equ1}) with Bayes rule,
\begin{equation}
    p(s | Y_T, Y_t, G) = \frac{p(Y_T| s, Y_t, G) p(s| Y_t, G) p(Y_t, G)}{p(Y_T, Y_t, G)},
\nonumber
\end{equation}
since $p(Y_T, Y_t, G)$ and $p(Y_t, G)$ are independent of $s$ (it can be seen as a propagation rule that follows the network topology) and can be disregarded. Thus a conditional probability can be obtained,
\begin{equation}
     p(s | Y_T, Y_t, G) \propto p(Y_T| s, Y_t, G) p(s| Y_t, G),
\nonumber
\end{equation}
where $p(s | Y_t, G)$ is the conditional probability distribution. The task then shifts to employing the Maximum A Posteriori (MAP) approximation to estimate the optimal diffusion source $\hat{s}$ by maximizing the following probability,
\begin{equation}
\begin{split}
    \hat{s} &= \mathop{\arg\max}\limits_{s} p(Y_T| s, Y_t, G) p(s| Y_t, G) \\
            &= \mathop{\arg\max}\limits_{s} \big( \log p(Y_T| s, Y_t, G) + \log p(s| Y_t, G) \big).
\label{equ:argmax-problem}
\end{split}
\end{equation}
\textbf{Theorem 3.1 (Reformulation of conditional distribution).}
\textit{Since the conditional distribution $p(s| Y_t, G)$ is not exactly applicable, we introduce a continuous latent variable $z$, and the empirical joint distribution $ q(s, Y_t, G, z)$. The objective becomes to minimize the divergence between $p(s, Y_t, G, z)$ and $q(s, Y_t, G, z)$,}
\begin{equation}
    \min Div \Big(q(s, Y_t, G, z || p(s, Y_t, G, z)) \Big).
\end{equation}
\begin{tcolorbox}[colback=gray!10, colframe=gray!10, boxrule=0pt, rounded corners, breakable, left=0pt, right=0pt]
\textit{Proof.} By the law of total probability, the conditional probability $(s|Y_t, G)$ can be written as,
\begin{equation}
\nonumber
    p(s|Y_t, G) = \int p(s, z|Y_t, G) dz.
\label{equ: map}
\end{equation}
Using the definition of conditional probability, the joint distribution $ p(s, z|Y_t, G) $ can be decomposed as,
\begin{equation}
\nonumber
	p(s, z | Y_t, G) = p(s | Y_t, G, z)p(z | Y_t, G).
\end{equation}
Substituting this into the total probability formula, we obtain,
\begin{equation}
\nonumber
	p(s|Y_t, G) = \int p(s | Y_t, G, z)p(z | Y_t, G) dz
\end{equation}
Given the propagation dynamics $Y_t$, and the associated graph $G$, $p(Y_t, G) = 1$, indicating that $ Y_t $ and $ G $ are fixed known conditions. Thus, the joint distribution $ p(s, Y_t, G, z)$ can be expressed as,
\begin{equation}
\nonumber
	p(s, Y_t, G, z) = p(s, z | Y_t, G)p(Y_t, G),
\end{equation}
Since $p(Y_t, G) = 1$, we have $ p(s, Y_t, G, z) = p(s | Y_t, G, z)p(z | Y_t, G)$. To approximate the true distribution $ p(s, Y_t, G, z)$, we introduce a varitional family,
\begin{equation}
\nonumber
	q(s, Y_t, G, z) = q(z|s, Y_t, G)\widetilde{q}(s, Y_t, G),
\end{equation}
Where $\widetilde{q}(s, Y_t, G)$ denotes the empirical distribution from the data. To align the model distribution with the empirical distribution, we minimize the divergence between the two joint distributions, yielding the objective,
\begin{equation}
\nonumber
    \min Div \Big(q(s, Y_t, G, z || p(s, Y_t, G, z)) \Big).
\end{equation}
\end{tcolorbox}
\textbf{Theorem 3.2 (Variational reformulation via ELBO).} \textit{Given $p(s, Y_t, G, z)$ and $q(s, Y_t, G, z)$, we introduce the Evidence Lower Bound \textup{(\textbf{ELBO})}, $\textup{\textbf{ELBO}} = \mathbb{E}_{z} \ \ln \ p(s | Y_t, G, z) - KL (q(z|s, Y_t, G) || p(z|Y_t,G) )$, then minimizing the kullback-Leibler (KL) divergence between $p$ and $q$ is equivalent to minimizing the negative \textup{\textbf{ELBO}},}
\begin{equation}
	\min KL(q||p) \Longleftrightarrow \min -\textbf{ELBO}. 
\end{equation}
\begin{tcolorbox}[colback=gray!10, colframe=gray!10, boxrule=0pt, rounded corners, breakable, left=0pt, right=0pt]
\textit{Proof.} The KL divergence between $q(s, Y_t, G, z)$ and $p(s, Y_t, G, z)$ can be calculated as,
\begin{equation}
\nonumber
\begin{split}
&KL\big( q(s, Y_t, G, z) || p(s, Y_t, G, z) \big) \\
&= \mathbb{E}_{ q(s, Y_t, G, z)}\Big[\ln\frac{ q(s, Y_t, G, z)}{ p(s, Y_t, G, z)}\Big].
\end{split}
\end{equation}
Given $ q(s, Y_t, G, z) = q(z|s, Y_t, G)\widetilde{q}(s, Y_t, G)$ and $ p(s, Y_t, G, z) = p(s | Y_t, G, z)p(z | Y_t, G)$, we obtain,
\begin{equation}
\nonumber
\begin{split}
   & KL(q(s, Y_t, G, z || p(s, Y_t, G, z)) \\
   & = \mathbb{E}_{s\sim \widetilde{q}(s, Y_t, G)}\ln \widetilde{q}(s, Y_t, G) + \mathbb{E}_{s\sim \widetilde{q}(s, Y_t, G)} \\
   &\Big( \mathbb{E}_{z\sim q(z| s, Y_t, G)} -\ln p(s | Y_t, G, z) \\
   & + KL(q(z|s, Y_t, G) || p(z|Y_t, G))  \Big).
\end{split}
\label{equ:KL(p||q)}
\end{equation}
For simplicity, we omit the subscript of $\mathbb{E}_{s\sim \widetilde{q}(s, Y_t)}$ and $\mathbb{E}_{z\sim q(z| s, Y_t, G)}$ as $\mathbb{E}_{s}$ and $\mathbb{E}_{z}$. Rearranging terms yields,
\begin{equation} \small
\nonumber
\begin{split} 
    & KL \Big(q(s, Y_t, G, z)||p(s, Y_t, G, z) \Big) - \mathbb{E}_{s} \ln \widetilde{q}(s, Y_t, G) \\
    & = \mathbb{E}_{s} \Big(- \mathbb{E}_{z} \ln  p(s | Y_t, G, z) + KL(q(z|s, Y_t, G) || p(z|Y_t,G)) \Big).
\end{split}
\label{equ:KL(p||q) - E}
\end{equation}
We introduce Evidence Lower BOund (\textbf{ELBO}), $ \textbf{ELBO} =  \mathbb{E}_{z} \ln p(s | Y_t, G, z) - KL \Big(q(z|s, Y_t, G) || p(z|Y_t,G) \Big)$. The above Equation can be transformed into,
\begin{equation}
\nonumber
\begin{split}
    &\mathbb{E}_{s} \ln \widetilde{q}(s, Y_t, G) \\
    &= \mathbb{E}_{s} ELBO + KL \Big(q(s, Y_t, G, z)||p(s, Y_t, G, z) \Big),
\end{split}
\end{equation}
where $\mathbb{E}_{s} \ln \widetilde{q}(s, Y_t)$ is a constant. This means that minimizing $KL(q(s, Y_t, G, z)||p(s, Y_t, G, z))$ is equivalent to minimizing negative \textbf{ELBO},
\begin{equation}
\nonumber
    \min -\mathbb{E}_{z} \ln p(s | Y_t, G, z) + KL \Big(q(z|s, Y_t, G) || p(z|Y_t,G) \Big).
\label{equ:min-ELBO}
\end{equation}

\end{tcolorbox}

\begin{figure}
    \centering
    \includegraphics[width=0.45\textwidth]{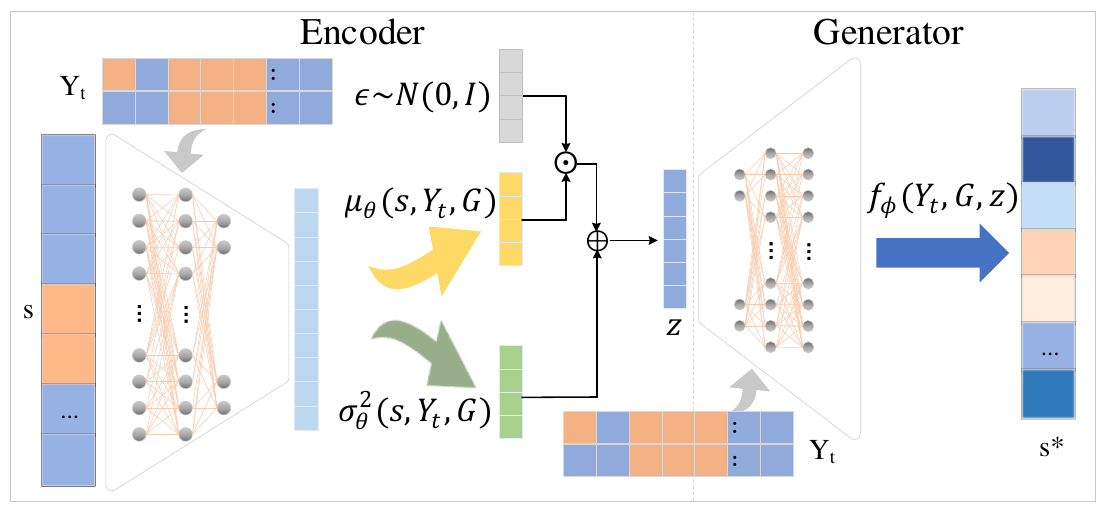}
    \caption{An illustration of the VAE variant.}
    \label{fig:CVAE-VAE}
\end{figure}

To accomplish this task, deep generative models provide an effective solution that models the prior distribution through a deterministic transformation \cite{hegde2018algorithmic}. Variational Auto-Encoders (VAEs) \cite{kingma2013auto}, as typical generative methods, have been widely used to learn the generative priors in many inverse problems \cite{peng2021generating, duan2023qarv}. They integrate a differentiable encoder network with a generative network, enabling the learning of complex data distributions. To mitigate the issue of blurred generated samples, often a consequence of noise introduction, CVAE \cite{sohn2015learning} employs additional conditionality for conditional generation. With this idea, we design a VAE variant that uses propagation dynamics as encoding and generation conditions. As illustrated in Fig. \ref{fig:CVAE-VAE}, there are four types of variables: input variables $\boldsymbol{s}$, latent variables $\boldsymbol{z}$, conditional variables $\boldsymbol{Y_t}$, and output variables $\boldsymbol{\hat{s}}$, the main difference between this variant and VAE is whether there are additional constraints $\boldsymbol{Y_t}$ (dotted arrows) in the encoder and generator. 
Therefore, we construct the likelihood $q_\theta(z|s, Y_t, G)$ of the encoder $f_\theta$ parameterized by $\theta$, and the likelihood $p_\phi(s|Y_t, G, z)$ of the generator $f_\phi$ parameterized by $\phi$.

We model the latent variable $z$ as following a multivariate normal distribution, $z \sim N(0, I)$, to enable efficient sampling. The encoder parameterized the posterior distribution by constructing means $\mu_\theta$ and variance $\sigma_\theta^2$ as functions of $s, Y_t, G$,
\begin{equation}
\begin{split}
    q_\theta(z|s, Y_t, G) &= N(\mu_\theta(s, Y_t, G), \sigma_\theta^2(s, Y_t, G)) \\
                            &= \mu_\theta(s, Y_t, G) + \sigma_\theta^2(s, Y_t, G) \odot \epsilon,
\end{split}
\end{equation}
where $\mu_\theta(s, Y_t, G)$ and $\sigma_\theta^2(s, Y_t, G)$ are mean and variance of neural networks with input $s, Y_t, G$, $\epsilon \sim N(0,I)$ is an auxiliary variable, $\odot$ is element-wise multiplication. In this way, the sampling process only involves linear operations, which are differentiable. The KL divergence in ELBO can be calculated as,
\begin{equation}
\begin{aligned}
    & \mathcal{L}_{KL} = KL \Big(q_\theta(z|s, Y_t, G) || p(z | Y_t, G) \Big) \\
    & = \frac{1}{2} \Big(-\ln\sigma_\theta^2(s, Y_t, G) + \sigma_\theta^2(s, Y_t, G) + \mu_\theta^2(s, Y_t, G) - 1 \Big).
\end{aligned}
\label{equ:L_KL}
\end{equation}
The derivative process is provided in the APPENDIX \ref{appendix}. The first term in \textbf{ELBO} corresponds to the generator of the model and can be viewed as the reconstruction of propagation source $s$ based on the latent variable $z$. Since $s$ is binary data (either infected or not) and fits the Bernoulli distribution, we can derive the likelihood function for optimizing $\phi$ as,
\begin{equation}
    \mathcal{L}_{re} = -\ln \sum_{j=1}^{N} \prod_{i=1}^{n} \Big[ f_\phi(Y_t^j, G, z^j)_i^{x_i^j} \cdot (1 - f_\phi(Y_t^j, G, z^j)_i)^{1-x_i^j}\Big],
\label{equ:L_re}
\end{equation}
where $N$ and $n$ are the number of training samples and nodes in the network, $f_{\phi}(\cdot)_i$ is the infected probability value of node $i$ generated by the generator.

\subsection{Forward Propagation}
The subsequent objective involves predicting the diffusion outcome $Y_T^\triangle$ conditioned on the network topology $G$, the propagation dynamics $Y_t$, and candidate sources $s^*$, as depicted by $p(Y_T| s, Y_t, G)$ in Equation (\ref{equ:argmax-problem}). 
While individual user actions in information diffusion are typically observed as discrete binary events (e.g., a binary 'retweet' or 'no retweet' action), the underlying social influence and the probability of a node transitioning between states evolve continuously over time.
To capture this latent continuous evolution, we adopt Graph Neural ODEs \cite{chen2018neural, poli2019graph} to extend discrete GNNs to continuous scenarios, denoted by $p_\psi(Y_T| s^*, Y_t, G)$, where the model depth corresponds to a continuous temporal notion \cite{xhonneux2020continuous}. Unlike the conventional discrete GNN layers, this continuous formulation provides a natural alignment with fundamental diffusion equations \cite{chamberlain2021grand}. The node state of the network given the terminal time step $T$ can be solved by a designated ODE solver,
\begin{equation}
\begin{split}
    H_T &= ODESolver(\frac{d H_t}{d t}, s^*, T) \\
                  &= s^* + \int_0^T f_\psi(H_t, t) dt,
\end{split}
\end{equation}
where $f_\psi$ denotes the Graph Neural ODEs function and parameterizes the continuous dynamics of node states, $H_t$ is the node states at time $t$. The neural ODEs consider the parameter updating in neural networks as solving an initial value problem. From a numerical analysis perspective, traditional discrete layer architectures implicitly approximate this continuous system through Euler discretization with fixed step sizes. For higher precision, we employ a typical Runge-Kutta-4 solver \cite{runge1895numerische} with fixed step size $\delta$, it would iteratively update the value $H_t$ with,
\begin{equation}
\begin{split}
k_i &= \left\{
\begin{aligned}
    \ &f_\psi(H_t)  &i=1, \\
    \ &f_\psi(H_t + \delta \cdot k_1 /2) \ &i=2, \\
    \ &f_\psi(H_t + \delta \cdot k_2 /2) \ &i=3, \\
    \ &f_\psi(H_t + \delta k_3) \ &i=4,
\end{aligned}
\right. \\
H_{(t+\delta)} &= H_t + \delta \cdot (k_1 + 2k_2 + 2k_3 + k_4) / 6.
\end{split}
\end{equation}

In this way, the output $H_T \in \mathbb{R}^{n \times d}$ of Graph Neural ODEs is equivalent to the result of multiple rounds of information dissemination, and we map it into the probability distribution through multi-layer perceptron (MLP),
\begin{equation}
    Y_T^\triangle = MLP(H_T),
\end{equation}
where $Y_T^\triangle \in \mathbb{R}^n$ is the continuous infection probability of nodes in the range $[0,1]$. The mean square error (MSE) can be used to update the parameter $\psi$ in this module,
\begin{equation}
    \mathcal{L}_{fp} = \frac{1}{n}\sum_{i=1}^{n}\big(Y_T(i) - Y_T^{\triangle}(i)\big)^2,
\end{equation}
where $n$ is the number of nodes, $Y_T(i)$ is the true state of the node $i$. Therefore, the log-likelihood of Equation (\ref{equ:argmax-problem}) in the training phase is defined as,
\begin{equation}
    \mathcal{L}_{training}(\theta,\phi,\psi) = \mathcal{L}_{KL}
    + \mathcal{L}_{re} + \mathcal{L}_{fp} + \mathcal{L}_{reg}(\theta,\phi,\psi),
\label{equ:L_training}
\end{equation}
where $\mathcal{L}_{reg}$ is an L2-norm regularization term that encompasses all parameters, commonly used to avoid the issue of overfitting. The training process is summarized in Algorithm \ref{algo:training_process} and Fig. \ref{fig:framework}.
\begin{algorithm}
    \caption{Training Phase}
    \begin{algorithmic}[1]
        \Require {source vector: $s \in \mathbb{R}^n$, propagation dynamics: $Y_t \in \mathbb{R}^{(T-1) \times n}$, propagation result: $Y_T \in \mathbb{R}^n$, adjacency matrix of graph $G$: $A \in \mathbb{R}^{n \times n}$}
        \Ensure {update $\theta$, $\phi$, $\psi$}
        \For{each $epoch$ in $num\_epochs$}
        \State  $\mu, \sigma \gets f_\theta(s, Y_t, A)$ \Comment {Encoder $f_\theta$}
        \State $z \gets \mu + \sigma \odot \epsilon$ \Comment {Reparameterization trick}
        \State $ s^* \gets f_\phi(Y_t, A, z) $ \Comment{Generator $f_\phi$}
        \State $H_T \gets f_\psi(s^*, A, Y_t)$ \Comment{Graph Neural ODEs}
        \State $Y_T^\triangle \gets MLP(H_T)$ \Comment{Mapping function}
        \State Calculate $\mathcal{L}_{training}$ \Comment{Equation (\ref{equ:L_training})}
        \State Minimize $\mathcal{L}_{training}$ and update parameters $\theta, \phi, \psi$
        \EndFor
    \end{algorithmic}
\label{algo:training_process}
\end{algorithm}

\subsection{Source Inference}
After the training process, the candidate source distribution of $p(s^*)$ is generated from latent variable $z$, which is obtained by encoder $f_\theta(s, Y_t, G)$. To address the unobservability of the ground-truth source vector $s$ during the inference phase, we implement a matching strategy to provide a robust prior for the latent variable $z$. This strategy leverages the archived knowledge within the training set to identify historical diffusion patterns that closely resemble the current observation. 

Formally, let $Y_T(test) \in \mathbb{R}^n$ be the observed final propagation state. We calculate the Wasserstein distance between $Y_T(test)$ and the propagation results of all training blocks $Y_T(train) = \{ Y_T^b(train)\}_{b=1}^B$, where $B$ is the number of training batches,
\begin{equation}
\begin{aligned}
&W(Y_T(test), Y_T^b(train)) \\
&=\inf_{\gamma \in \Pi(Y_T(test), Y_T^b(train))} \mathbb{E}_{(x,y) \sim \gamma} [||x-y||].
\end{aligned}
\end{equation}
The strategy identifies the index that exhibits the minimal distributional divergence,
\begin{equation}
s_b = \mathop{\arg\min}\limits_{b \in \{1,2,\cdots,B\}} W(Y_T(test), Y_T^b(train)),
\end{equation}
where $s_b$ is the propagation source corresponding to the batch with the highest similarity in the training set. Therefore, the latent variable can be calculated as,
\begin{equation}
    p(\Bar{z}) = \frac{1}{|s_b|} \sum_{i=1}^{|s_b|} f_\theta (z|s_i, Y_t, G),
\label{equ:z_bar}
\end{equation}
where $Y_t$ is the propagation dynamics of the test set, which is equivalent to the conditional signal of the encoding process. It is important to note that the space complexity of the matching strategy is $\mathcal{O}(M\cdot |V|)$, where $M$ denotes the number of training samples. This storage is independent of the model’s trainable parameters and serves as an auxiliary offline component. Therefore, it does not increase the model's parameter scale or its internal memory requirements. The test process can be defined as,
\begin{equation}
    \hat{s} = \mathop{\arg\max}\limits_{s^*} {p_\psi(Y_T | s^*, Y_t, G) p(s^*)}.
\nonumber
\end{equation}
Since $Y_t$ and $G$ are known, to obtain the distribution $p(s^*)$, we calculate its marginal distribution through latent variable $\Bar{z}$, which is calculated by Equation (\ref{equ:z_bar}). Hence, the above formula can be transformed into,
\begin{equation}\small
\begin{aligned}
    &\hat{s} = \mathop{\arg\max}\limits_{s^*} {p_\psi(Y_T | s^*, Y_t, G) \sum_{\Bar{z}} p_\phi(s^*|\Bar{z}, Y_t, G) p(\Bar{z})} \\
    &=\mathop{\arg\max}\limits_{s^*} {p_\psi(Y_T | s^*, Y_t, G) \sum_{z} \sum_{s_b} p_\phi(s^*|z, Y_t, G) q_\theta(z|s_b, Y_t, G)},
\end{aligned}
\end{equation}
where $\psi, \phi$ and $\theta$ are parameters determined during the training process. The log likelihood function of the above formula can be expressed as,
\begin{equation}
\begin{split}
    &\min\limits_{s^*} \big [-\log p_\psi(Y_T | s^*, Y_t, G) \\
    &- \log \sum_{z} \sum_{s_b} p_\phi(s^*|z, Y_t, G) q_\theta(z|s_b, Y_t, G) \big].
    \end{split}
\end{equation}
The first term $p_\psi(Y_T | s^*, Y_t, G)$ is the forward propagation module that outputs the infection probability of nodes. As in the training process, we use mean square error (MSE) to measure the inference error between true results $Y_T$ and predicted results $Y_T^\triangle$. The second term can be transformed into a reconstruction form similar to Equation (\ref{equ:L_re}). Therefore,
\begin{equation}
\begin{aligned}
    \mathcal{L}&_{infer} (s^*) = log \frac{1}{n}\sum_{i=1}^{n}\big(Y_T(i) -  f_\psi(Y_T(i) | s^*, Y_t, G) \big)^2 \\
    &- log \sum_{j=1}^{B} \prod_{i=1}^{n} \Big[ f_\phi(Y_t, G, z^j)_i^{x_i^j} \cdot (1 - f_\phi(Y_t, G, z^j)_i)^{1-x_i^j}  \Big],
\label{equ:L_infer}
\end{aligned}
\end{equation}
where $B$ denotes the batch block size, which is most similar to the propagation result $Y_T$ in the training set, $n$ is the number of nodes in the network. $z^j$ is encoded through $f_\theta(s_j, Y_t, G)$. The general inference process is illustrated in Algorithm \ref{algo:inference} and Fig. \ref{fig:framework}.

\begin{algorithm}
    \caption{Inference Phase}
    \begin{algorithmic}[1]
        \Require{propagation dynamics: $Y_t \in \mathbb{R}^{(T-1) \times n}$, propagation result: $Y_T \in \mathbb{R}^n$, adjacency matrix of graph $G$: $A \in \mathbb{R}^{n \times n}$}, parameters: $\theta, \phi, \psi$, learning rate: $\alpha$
        \Ensure{predicted source vector: $\hat{s} \in \mathbb{R}^n$}
        \State $s_b \gets Div(Y_T(train), Y_T(test))$ \Comment{Identify similar batch}
        \State $Sampling \ latent \ variables \ \Bar{z}$ \Comment{Equation(\ref{equ:z_bar})}
        \State $\widetilde{s}^* \gets f_\phi(Y_t, A, \Bar{z})$ \Comment{Generate an initial source $\widetilde{s}^*$}
        \For{each $epoch$ in $num\_epochs$}
        \State $H_T \gets f_\psi(\widetilde{s}^*, A, Y_t)$ \Comment{Graph Neural ODEs}
        \State $Y_T^\triangle \gets MLP(H_T)$ \Comment{Mapping function}
        \State Calculate $\mathcal{L}_{infer}$ \Comment{Equation (\ref{equ:L_infer})}
        \State $\widetilde{s}^* \gets \alpha \cdot \nabla \mathcal{L}_{infer}$ \Comment{Update initial source vector $\widetilde{s}^*$}
        \EndFor
        \State $\hat{s} \gets \widetilde{s}^*$ \Comment{Output the final source vector}
    \end{algorithmic}
\label{algo:inference}
\end{algorithm}

\section{Experiment}\label{sec:experiment}
In this section, we conduct experiments on both synthetic and real-world diffusion datasets to comprehensively validate the effectiveness of the proposed framework. Prior to analyzing the results, we introduce the datasets, implementation details, evaluation metrics, and baselines.

\begin{table}[htbp]\scriptsize
  \centering
  \caption{Statistics of The Datasets.}
  \setlength{\tabcolsep}{1mm}{
    \begin{tabular}{c|ccccccc}
    \toprule
          & Jazz  & NS & CM & Cora-ML & BA & Twitter & Douban  \\
    \midrule
    Directed/Undirected & Di & Un & Di & Un & Un & Un & Un \\
    Nodes & 198   & 1,461  & 1,899  & 2,708  & 1,000 & 12,627 & 24,688 \\
    Edges & 2,742  & 2,742  & 20,296 & 5,278  & 1,997 & 309,631 & 379,155 \\
    Avg. degree & 27.696 & 3.753 & 21.375 & 3.898 & 3.994 & 49.042 & 30.715 \\
    Clustering Coef. & 0.308 & 0.693 & 0.087 & 0.240  & 0.011 & 0.296 & 0.112 \\
    Density & 0.070  & 0.002 & 0.005 & 0.001 & 0.004 & 0.004 & 0.001 \\
    Source Ratio & 0.05 & 0.07 & 0.10 & 0.10 & 0.05 & - & - \\
    \bottomrule
    \end{tabular}}%
  \label{tab:datasets}%
\end{table}%

\subsection{Datasets}
We generate synthetic diffusion datasets across five networks: Jazz \cite{rossi2015network}, Network Science \cite{rossi2015network}, CollegeMsg \cite{panzarasa2009patterns}, Cora-ML \cite{rossi2015network}, and BA networks \cite{barabasi1999emergence}. Additionally, we incorporate two real-world diffusion datasets, Twitter \cite{hodas2014simple} and Douban \cite{zhong2012comsoc}, to further validate the performance in the real scenario. These datasets vary in domain, scale, and density, as outlined in the Table \ref{tab:datasets}.

In the experiments, we randomly select 5\%, 7\%, and 10\% of the nodes as sources to generate synthetic cascades. Current research on source localization mainly focuses on inference in simple contagion scenarios (i.e., SI, SIR, IC). To comprehensively evaluate the performance of the proposed framework, we use SI \cite{allen1994some}, SIR \cite{allen1994some}, and GLT \cite{ran2020generalized} models, which correspond to simple (SI, SIR) and complex (GLT) contagion processes. 
Each edge is assigned a random weight $\omega \in (0.2,0.8)$, representing the pairwise transition probability. This randomness in the cascade generation process allows the node contagion to be significantly affected by its intrinsic properties and the random numbers (similar external factors). Such processing is closer to the complexity of dissemination phenomena in the real world and enhances the model's generalization ability. For the GLT model, each node $i$ is associated with an activation threshold $\tau = k^{degree_i}$, defined as the minimum weighted influence required from its neighbors to trigger propagation. In order to ensure early-stage diffusion viability under GLT, seed nodes are exclusively sampled from the top 50\% of node IDs. The simulation unfolds in continuous time steps, and we discard cascades where the final infected nodes fall below 40\% of the total network size. Valid cascades are partitioned into training and testing subsets, with an 80\% and 20\% ratio, respectively.

For both Twitter \cite{hodas2014simple} and Douban \cite{zhong2012comsoc} datasets, we design two experimental settings. In each cascade sequence, we consider (i) selecting the first 20 nodes as a fixed number of sources, and (ii) selecting the top 10\% of nodes as the unfixed number of sources. Since real-world networks typically exhibit large-scale global structures while actual information propagation is often confined to localized subnetworks, we project each information cascade onto its corresponding subgraph structure derived from the global network. This subgraph encompasses all participating nodes in the cascade and their associated edges. Consequently, the model is required to output tailored predictive distributions for different subgraph structures. To mitigate prediction bias, we randomly shuffle the node IDs in each cascade to ensure that the positions of sources in the prediction vector are not fixed.

Furthermore, we partition the cascade, excluding the source nodes, into $T$ temporally ordered snapshots, where each snapshot contains an equal number of newly infected nodes. The final snapshot represents the propagation result. 

\subsection{Experiment Setup}
\subsubsection{Implementation Details}
For the proposed model, the encoder $f_\theta$ and the generator $f_\phi$ utilize two-layer MLPs with non-linear activation. To incorporate the propagation dynamics $Y_t$ into the encoding and generation process, we concatenate the network state vector $Y_{t_i}$ corresponding to each snapshot $i$ sequentially, making it part of the encoder and generator input. In the forward propagation model, we set the hidden dimensions of GNN $f_\psi$ as 128. 
The three-layer MLP following Graph Neural ODEs has 128 hidden units in each layer. The learning rates are set to 0.03 for the training phase and 0.001 for the test phase. The number of training epochs is set to 100 in Algorithm \ref{algo:training_process} and the number of epochs for initial source revision is set to 15 in Algorithm \ref{algo:inference}. 
For baselines, we adopt the default parameter settings provided in the open-source implementations. We release the code at \url{https://github.com/MrYansong/PDSL}.

\subsubsection{Evaluation Metrics}
Source detection can be viewed as a multilabel classification problem, where each node corresponds to a category, and each label is binary $\{0, 1\}$ across $|V|$ categories. We use four metrics to evaluate performance: Macro-Precision, the average ratio of true positive predictions to the total number of positive predictions across all categories, Macro-Recall, the average ratio of true positive predictions to the total number of true positive labels across all categories, Macro-F1, the average F1 score across all categories, and Accuracy, the ratio of correct predictions to all categories. Macro-scores represent the average of scores for each class in a multilabel classification problem,
\begin{equation}
    \text{scores}_{\text{macro}} = \frac{1}{n} \sum\limits_{i=1}^n \text{score}_i,
\nonumber
\end{equation}
where $n$ is the number of categories. These metrics provide a balanced view of the performance across all classes in a classification problem, which is especially useful when dealing with imbalanced datasets. Each case undergoes 10 independent runs for statistical robustness.

\begin{table}[!htbp]\setlength{\tabcolsep}{0.8mm}\scriptsize
  \centering
  \caption{Comparison of Input Requirements and Assumptions across Different Methods.}
    \begin{tabular}{c|ccccccccc}
    \toprule
          & LPSI  & Netsleuth & OJC   & GCNSI & IVGD  & SLVAE & BOSouL & SDSA  & Ours \\
    \midrule
    Diffusion & \ding{51} & \ding{51} & \ding{51} & \ding{51} & \ding{51} & \ding{51} & \ding{51} & \ding{51} & \ding{55} \\
    Seed Count & \ding{51} & \ding{51} & \ding{51} & \ding{55} & \ding{55} & \ding{55} & \ding{51} & \ding{51} & \ding{55} \\
    Temporal & \ding{55} & \ding{55} & \ding{55} & \ding{55} & \ding{55} & \ding{55} & \ding{55} & Fine & Coarse \\
    \bottomrule
    \end{tabular}%
  \label{tab:features}%
\end{table}%

\subsubsection{Baselines}
To evaluate the performance of our proposed model against other state-of-the-art source detection models. We use eight baselines:
\begin{itemize}[leftmargin=1em]
\item Netsleuth \cite{prakash2012spotting} employs the minimum description length principle to identify the best set of source nodes.
\item LPSI \cite{wang2017multiple} assigns labels to the partially infected network and propagates these labels on a weighted network until convergence, identifying nodes that meet specific conditions as sources.
\item OJC \cite{zhu2017catch} finds sources based on the Jordan cover that ensure all observed infected nodes with the minimal radius.
\item GCNSI \cite{dong2019multiple} utilizes neighbors' information and spectral domain convolution to locate multiple sources without relying on predefined diffusion mechanisms.
\item IVGD \cite{wang2022invertible} uses a series of linearization techniques, including invertible graph residual nets, error compensation modules, and unrolled optimization, to locate sources from an inverse problem perspective.
\item SLVAE \cite{ling2022source} is the first method to leverage deep graph generative models, VAEs, to tackle the source detection.
\item BOSouL \cite{zhang2024multiple} is a simulation based method that adopts Bayesian optimization to reveal the relationship between the node set and the observation.
\item SDSA \cite{cheng2025efficient} is a sensor based method that combines greedy sensor deployment with Bayesian-based estimation.
\end{itemize}

We analyze the input requirements and assumptions of all baselines alongside our proposed approach. As summarized in Table \ref{tab:features}, most existing methods (e.g., LPSI, NetSleuth, OJC) rely heavily on prior knowledge, such as the specific diffusion mechanism or the exact seed count. In contrast, our method operates in a mechanism-agnostic manner. Furthermore, unlike SDSA which demands fine-grained temporal data, our approach utilizes coarse-grained snapshots, making it more applicable to real-world scenarios where continuous monitoring is impractical. Notably, empirical results in the ablation study indicate that our model achieves comparable or superior performance even in the absence of temporal information.

\begin{table*}[!htb]\setlength{\tabcolsep}{0.9mm}
    \caption{Performance Comparison under SI Diffusion Pattern.}
    \centering
    \setlength{\tabcolsep}{0.9mm}{
    \begin{tabular}{c|c c c c|c c c c | c c c c | c c c c | c c c c}
    \toprule
   \multicolumn{1}{c|}{} & \multicolumn{4}{c|}{Jazz} &\multicolumn{4}{c|}{Network science} &\multicolumn{4}{c|}{CollegeMsg} &\multicolumn{4}{c|}{Cora-ML} &\multicolumn{4}{c}{BA} \\
    \midrule
    \multicolumn{1}{c|}{Method} & Pr & Re & F1 & Acc & Pr & Re & F1 & Acc & Pr & Re & F1 & Acc & Pr & Re & F1 & Acc & Pr & Re & F1 & Acc \\
    \midrule
    LPSI & 0.529 & 0.651 & 0.415 & 0.539 & 0.586 & 0.778 & 0.579 & 0.753 & 0.602 & 0.788 & 0.557 & 0.657 & 0.617 & 0.739 & 0.565 & 0.709 & 0.543 & 0.720 & 0.450 & 0.588 \\
    Netsleuth & 0.529 & 0.542 & 0.533 & 0.890  & 0.583 & 0.620  & 0.586 & 0.874 & 0.596 & 0.606 & 0.600   & 0.856 & 0.591 & 0.621 & 0.573 & 0.848 & 0.529 & 0.545 & 0.533 & 0.887 \\
    OJC   & 0.553 & 0.757 & 0.473 & 0.623 & 0.544 & 0.500   & 0.484 & 0.921 & 0.622 & 0.501 & 0.478 & 0.865 & 0.545 & 0.500   & 0.475 & 0.856 & 0.538 & 0.512 & 0.503 & 0.919 \\
    GCNSI & 0.509 & 0.545 & 0.469 & 0.733 & 0.575 & 0.501 & 0.484 & 0.921 & 0.584 & 0.603 & 0.592 & 0.844 & 0.552 & 0.649 & 0.437 & 0.503 & 0.531 & 0.510   & 0.516 & 0.915 \\
    IVGD  & 0.545 & \textbf{0.762} & 0.438 & 0.549 & 0.595 & \textbf{0.843} & 0.567 & 0.708 & 0.606 & \textbf{0.805} & 0.554 & 0.647 & 0.607 & \textbf{0.838} & 0.573 & 0.709 & 0.550  & \textbf{0.762} & 0.435 & 0.549 \\
    SLVAE & 0.517 & 0.522 & 0.508 & 0.913 & 0.510  & 0.512 & 0.510  & 0.864 & 0.619 & 0.520  & 0.517 & 0.893 & 0.576 & 0.526 & 0.521 & 0.884 & 0.480  & 0.505 & 0.491 & 0.923 \\
    BOSouL & 0.553 & 0.544 & 0.547 & 0.929 & 0.576 & 0.545 & 0.560  & 0.909 & 0.556 & 0.551 & 0.554 & 0.873 & 0.527 & 0.591 & 0.532 & 0.886 & 0.541 & 0.534 & 0.521 & 0.894 \\
    SDSA  & 0.456 & 0.529 & 0.510  & 0.915 & 0.539 & 0.441 & 0.457 & 0.896 & 0.606 & 0.543 & 0.494 & 0.815 & 0.561 & 0.545 & 0.553 & 0.878 & 0.461 & 0.455 & 0.441 & 0.916 \\
    Ours  & \textbf{0.768} & 0.614 & \textbf{0.644} & \textbf{0.946} & \textbf{0.656} & 0.572 & \textbf{0.593} & \textbf{0.933} & \textbf{0.691}  & 0.645 & \textbf{0.665} & \textbf{0.899} & \textbf{0.680}  & 0.672 & \textbf{0.654} & \textbf{0.901} & \textbf{0.574} & 0.523 & \textbf{0.536} & \textbf{0.930} \\
    \bottomrule
    \end{tabular}
    }
    \label{tab:performance-si}
\end{table*}

\begin{table*}[!htb]\setlength{\tabcolsep}{0.9mm}
    \caption{Performance Comparison under GLT Diffusion Pattern.}
    \centering
    \setlength{\tabcolsep}{0.9mm}{
    \begin{tabular}{c|c c c c|c c c c | c c c c | c c c c | c c c c}
    \toprule
   \multicolumn{1}{c|}{} & \multicolumn{4}{c|}{Jazz} &\multicolumn{4}{c|}{Network science} &\multicolumn{4}{c|}{CollegeMsg} &\multicolumn{4}{c|}{Cora-ML} &\multicolumn{4}{c}{BA} \\
    \midrule
   \multicolumn{1}{c|}{Method} & Pr & Re & F1 & Acc & Pr & Re & F1 & Acc & Pr & Re & F1 & Acc & Pr & Re & F1 & Acc & Pr & Re & F1 & Acc \\
    \midrule
    LPSI  & 0.547 & 0.723 & 0.472 & 0.623 & 0.597 & \textbf{0.847} & 0.571 & 0.714 & 0.607 & 0.797 & 0.572 & 0.678 & 0.605 & 0.723 & 0.575 & 0.774 & 0.545 & 0.733 & 0.450  & 0.585 \\
    Netsleuth & 0.560  & 0.597 & 0.567 & 0.901 & 0.588 & 0.627 & 0.561 & 0.875 & 0.580  & 0.588 & 0.583 & 0.843 & 0.601 & 0.623 & 0.573 & 0.849 & 0.525 & 0.539 & 0.528 & 0.885 \\
    OJC   & 0.557 & \textbf{0.786} & 0.464 & 0.593 & 0.622 & 0.501 & 0.485 & 0.931 & 0.609 & 0.501 & 0.475 & 0.885 & 0.561 & 0.500   & 0.475 & 0.883 & 0.538 & 0.501 & 0.489 & 0.932 \\
    GCNSI & 0.474 & 0.500   & 0.487 & 0.949 & 0.537 & 0.545 & 0.540  & 0.869 & 0.590  & 0.566 & 0.575 & 0.872 & 0.538 & 0.554 & 0.228 & 0.230  & 0.526 & 0.537 & 0.530  & 0.892 \\
    IVGD  & 0.558 & 0.784 & 0.463 & 0.591 & 0.596 & 0.846 & 0.571 & 0.714 & 0.611 & \textbf{0.816} & 0.569 & 0.667 & 0.616 & \textbf{0.837} & 0.581 & 0.742 & 0.551 & \textbf{0.768} & 0.442 & 0.560 \\
    SLVAE & 0.503 & 0.492 & 0.491 & 0.910  & 0.471 & 0.458 & 0.464 & 0.835 & 0.631 & 0.529 & 0.528 & 0.886 & 0.511 & 0.503 & 0.481 & 0.891 & 0.477 & 0.489 & 0.483 & 0.927 \\
    BOSouL & 0.577 & 0.583 & 0.577 & 0.934 & 0.575 & 0.551 & 0.559 & 0.862 & 0.529 & 0.550  & 0.533 & 0.872 & 0.521 & 0.563 & 0.531 & 0.878 & 0.513 & 0.522 & 0.515 & 0.925 \\
    SDSA  & 0.599 & 0.552 & 0.486 & 0.901 & 0.542 & 0.447 & 0.534 & 0.898 & 0.621 & 0.546 & 0.489 & 0.817 & 0.597 & 0.552 & 0.525 & 0.881 & 0.468 & 0.501 & 0.436 & 0.917 \\
    Ours  & \textbf{0.803} & 0.681 & \textbf{0.719} & \textbf{0.957} & \textbf{0.667} & 0.569 & \textbf{0.591} & \textbf{0.936} & \textbf{0.689} & 0.642 & \textbf{0.661} & \textbf{0.898} & \textbf{0.645} & 0.575 & \textbf{0.593} & \textbf{0.897} & \textbf{0.636} & 0.531 & \textbf{0.542} & \textbf{0.941} \\
    \bottomrule
    \end{tabular}
    }
    \label{tab:performance-glt}
\end{table*}

\begin{table*}[htb]\setlength{\tabcolsep}{0.9mm}
\caption{Performance Comparison under SIR Diffusion Pattern.}
  \centering
  \setlength{\tabcolsep}{0.9mm}{
    \begin{tabular}{c|c c c c|c c c c | c c c c | c c c c | c c c c}
    \toprule
          & \multicolumn{4}{c|}{Jazz}     & \multicolumn{4}{c|}{Network science} & \multicolumn{4}{c|}{CollegeMsg}  & \multicolumn{4}{c|}{Cora-ML}     & \multicolumn{4}{c}{BA} \\
    \midrule
    Method & \multicolumn{1}{c}{Pr} & \multicolumn{1}{c}{Re } & \multicolumn{1}{c}{F1} & \multicolumn{1}{c|}{Acc} & \multicolumn{1}{c}{Pr} & \multicolumn{1}{c}{Re } & \multicolumn{1}{c}{F1} & \multicolumn{1}{c|}{Acc} & \multicolumn{1}{c}{Pr} & \multicolumn{1}{c}{Re } & \multicolumn{1}{c}{F1} & \multicolumn{1}{c|}{Acc} & \multicolumn{1}{c}{Pr} & \multicolumn{1}{c}{Re } & \multicolumn{1}{c}{F1} & \multicolumn{1}{c|}{Acc} & \multicolumn{1}{c}{Pr} & \multicolumn{1}{c}{Re } & \multicolumn{1}{c}{F1} & \multicolumn{1}{c}{Acc} \\
    \midrule
    {LPSI} & 0.550  & \textbf{0.762} & 0.435 & 0.548 & 0.621 & \textbf{0.815} & 0.583 & 0.786 & 0.606 & 0.793 & 0.580  & 0.692 & 0.598 & 0.785 & 0.530  & 0.672 & 0.567 & \textbf{0.829} & 0.515 & 0.675 \\
    {Netsleuth} & 0.531 & 0.546 & 0.536 & 0.890  & 0.600   & 0.644 & 0.595 & 0.880  & 0.567 & 0.573 & 0.570  & 0.845 & 0.577 & 0.564 & 0.544 & 0.851 & 0.542 & 0.558 & 0.541 & 0.898 \\
    {OJC} & 0.536 & 0.685 & 0.416 & 0.533 & 0.575 & 0.501 & 0.484 & 0.931 & 0.562 & 0.500   & 0.476 & \textbf{0.905} & 0.565 & 0.500   & 0.478 & \textbf{0.911} & 0.538 & 0.501 & 0.489 & 0.949 \\
    {GCNSI} & 0.474 & 0.500   & 0.487 & \textbf{0.949} & 0.466 & 0.500   & 0.482 & 0.931 & 0.453 & 0.500   & 0.475 & \textbf{0.905} & 0.456 & 0.500   & 0.476 & \textbf{0.911} & 0.475 & 0.500   & 0.487 & \textbf{0.950} \\
    {IVGD} & 0.550  & 0.761 & 0.434 & 0.547 & 0.621 & 0.805 & 0.588 & 0.787 & 0.613 & \textbf{0.819} & 0.574 & 0.674 & 0.605 & \textbf{0.791} & 0.545 & 0.656 & 0.547 & 0.737 & 0.404 & 0.501 \\
    {SLVAE} & 0.595 & 0.558 & 0.550  & 0.909 & 0.566 & 0.558 & 0.558 & 0.889 & 0.568 & 0.525 & 0.515 & 0.868 & 0.549 & 0.510  & 0.495 & 0.867 & 0.513 & 0.500   & 0.493 & 0.911 \\
    {BOSouL} & 0.548 & 0.533 & 0.538 & 0.890  & 0.537 & 0.526 & 0.528 & 0.852 & 0.539 & 0.535 & 0.537 & 0.833 & 0.546 & 0.538 & 0.541 & 0.835 & 0.538 & 0.528 & 0.531 & 0.894 \\
    SDSA  & 0.481 & 0.506 & 0.435 & 0.911 & 0.596 & 0.532 & 0.539 & 0.902 & 0.584 & 0.511 & 0.454 & 0.834 & 0.556 & 0.517 & 0.450  & 0.857 & 0.475 & 0.415 & 0.500   & 0.913 \\
    Ours & \textbf{0.711} & 0.579 & \textbf{0.601} & 0.935 & \textbf{0.714} & 0.576 & \textbf{0.603} & \textbf{0.931} & \textbf{0.687} & 0.611 & \textbf{0.633} & 0.900 & \textbf{0.613} & 0.542 & \textbf{0.553} & 0.898 & \textbf{0.618} & 0.530  & \textbf{0.551} & 0.933 \\
    \bottomrule
    \end{tabular}%
    }
  \label{tab:performance-sir}%
\end{table*}%

\subsection{Performance Comparison}
The overall performance of our model and all baselines are shown in Table \ref{tab:performance-si}, Table \ref{tab:performance-glt}, and Table \ref{tab:performance-sir}. From the results, we have the following observations.

While PDSL exhibits slightly lower Recall on certain datasets compared to baselines such as LPSI and IVGD, it consistently achieves significantly higher Precision and F1-score. This indicates that traditional methods often adopt a more aggressive prediction strategy, identifying a larger set of candidate nodes to ensure coverage, which inherently boosts Recall but leads to a substantial increase in false positives. From a practical deployment perspective, investigation resources are typically finite in tasks such as rumor containment or epidemic tracing. A high-precision model ensures that intervention efforts are concentrated on the most probable sources, avoiding the high operational costs and resource dilution associated with verifying a large number of incorrectly identified non-source nodes.
In the Jazz network, PDSL achieves $0.644$, $0.719$, and $0.601$ F1 scores under SI, GLT, and SIR mechanisms, respectively, while the score of the suboptimal baseline is merely $0.547$, $0.577$, and $0.550$, improvements of about 18\%, 25\%, and 10\%, indicating its superior ability to identify sources in smaller datasets. 

Although deep learning methods have shown excellent performance in many fields, this does not mean that they are the best choice in source detection. We find that rule-based methods excel in the source detection problem (e.g., Netsleuth in several datasets) because they are specially optimized to meet the specific needs of the task. This targeted optimization may make them more advantageous than deep learning methods in specific application scenarios.
Since the recovery state introduced by the SIR propagation mechanism weakens or interrupts the continuous spread of information, the resulting cascade data exhibit greater complexity compared to those generated by the SI and GLT models. As shown in Table \ref{tab:performance-sir}, OJC and GCNSI entirely collapse, misclassifying all nodes as non-sources. While the performance of PDSL on this dataset declines or remains comparable relative to the other two datasets, it still outperforms the baselines, demonstrating strong robustness.

\begin{table*}[htb]\setlength{\tabcolsep}{0.9mm} 
  \centering
  \caption{Performance Comparison under Real-world Diffusion.}
    \begin{tabular}{c|cccc|cccc|cccc|cccc}
    \toprule
    \multicolumn{1}{r}{} & \multicolumn{8}{c}{Twitter}                                   & \multicolumn{8}{c}{Douban} \\
    \midrule
    \multicolumn{1}{l|}{Sources} & \multicolumn{4}{c|}{fixed} & \multicolumn{4}{c|}{unfixed} & \multicolumn{4}{c|}{fixed} & \multicolumn{4}{c}{unfixed} \\
    \midrule
          & Pr    & Re    & F1    & Acc   & Pr    & Re    & F1    & Acc   & Pr    & Re    & F1    & Acc   & Pr    & Re    & F1    & Acc \\
    \midrule
    LPSI  & 0.621 & \textbf{0.826} & 0.590  & 0.687 & 0.542 & \textbf{0.660} & 0.515 & 0.734 & 0.525 & \textbf{0.570} & 0.432 & 0.520  & 0.515 & 0.547 & 0.438 & 0.567 \\
    Netsleuth   & 0.577 & 0.609 & 0.557 & 0.819 & 0.533 & 0.572 & 0.533 & 0.829 & 0.521 & 0.530  & 0.521 & 0.791 & 0.514 & 0.523 & 0.512 & 0.798 \\
    OJC   & 0.540  & 0.516 & 0.492 & 0.884 & 0.543 & 0.534 & 0.504 & 0.902 & 0.520  & 0.516 & 0.503 & 0.817 & 0.522 & 0.505 & 0.496 & 0.884 \\
    GCNSI & 0.450  & 0.500   & 0.473 & 0.900   & 0.500   & 0.507 & 0.339 & 0.479 & 0.450  & 0.500   & 0.473 & 0.900   & 0.420 & 0.500   & 0.312 & 0.463 \\
    IVGD  & -     & -     & -     & -     & -     & -     & -     & -     & -     & -     & -     & -     & -     & -     & -     & - \\
    SLVAE & 0.512 & 0.516 & 0.500   & 0.813 & 0.491 & 0.510  & 0.499 & 0.901 & 0.493 & 0.501 & 0.480  & 0.776 & 0.512 & 0.518 & 0.504 & 0.873 \\
    BOSouL & 0.623 & 0.588 & 0.598 & 0.874 & 0.558 & 0.533 & 0.537 & 0.868 & 0.598 & 0.545 & 0.563 & 0.842 & 0.561 & 0.552 & 0.556 & 0.851 \\
    SDSA & -     & -     & -     & -     & -     & -     & -     & -     & -     & -     & -     & -     & -     & -     & -     & - \\
    Ours  & \textbf{0.786} & 0.673 & \textbf{0.708} & \textbf{0.914} & \textbf{0.687} & 0.585 & \textbf{0.604} & \textbf{0.932} & \textbf{0.645} & 0.568 & \textbf{0.581} & \textbf{0.876} & \textbf{0.637} & \textbf{0.564} & \textbf{0.575} & \textbf{0.895} \\
    \bottomrule
    \end{tabular}%
  \label{tab:performance-real-world}%
\end{table*}%

\begin{table*}[htb]\setlength{\tabcolsep}{1mm}
  \centering
  \caption{Performance Comparison Measured by Average Error Distance.}
    \begin{tabular}{c|ccc|ccc|ccc|ccc|ccc}
    \toprule
          & \multicolumn{3}{c|}{jazz} & \multicolumn{3}{c|}{Network science} & \multicolumn{3}{c|}{CollegeMsg} & \multicolumn{3}{c|}{Cora-ML} & \multicolumn{3}{c}{BA} \\
    \midrule
    Diffusion Mechanism & SI    & GLT   & SIR   & SI    & GLT   & SIR   & SI    & GLT   & SIR   & SI    & GLT   & SIR   & SI    & GLT   & SIR \\
    \midrule
    LPSI  & 1.372 & 1.366 & 1.484 & 3.174 & 3.106 & 3.102 & 1.753 & 1.741 & 1.352 & 2.846 & 2.736 & 2.845 & 2.449 & 2.72  & 2.533 \\
    Netsleuth   & 1.303 & 1.109 & 1.362 & 3.257 & 3.272 & 3.226 & 1.188 & 1.263 & 1.321 & 2.473 & 2.317 & 2.592 & 2.557 & 2.633 & 2.789 \\
    OJC   & 1.363 & 1.338 & 1.416 & 3.942 & 3.923 & 3.963 & 1.462 & 1.458 & 1.540 & 2.834 & 2.798 & 2.931 & 2.715 & 2.726 & 2.716 \\
    GCNSI & 1.534 & 1.596 & 2.221 & 3.609 & 3.675 & 3.792 & 1.906 & 1.903 & 2.003 & 2.511 & 2.782 & 2.651 & 2.479 & 2.802 & 2.587 \\
    IVGD  & 1.715 & 1.910 & 1.521 & 2.966 & 4.707 & 3.416 & 1.173 & 1.184 & 1.526 & \textbf{1.797} & 2.213 & 2.516 & 2.253 & 2.387 & 2.450  \\
    SLVAE & 1.565 & 1.918 & 1.559 & 3.811 & 3.124 & 3.188 & 1.684 & 1.851 & 1.501 & 2.198 & 3.772 & 3.615 & 2.729 & 2.731 & 2.568 \\
    BOSouL & 1.396 & 1.246 & 1.359 & 3.505 & 3.590 & 3.406 & 1.178 & \textbf{1.151} & 1.301 & 2.679 & 2.591 & 2.883 & 2.635 & 2.457 & 2.632 \\
    SDSA  & 1.303 & 1.431 & 1.45  & 3.205 & 3.161 & \textbf{2.918} & 1.513 & 1.471 & 1.368 & 2.037 & 2.102 & 2.415 & \textbf{2.104} & 2.309 & 2.519 \\
    Ours  & \textbf{1.229} & \textbf{1.078} & \textbf{1.334} & \textbf{2.869} & \textbf{2.962} & 2.947 & \textbf{0.961} & 1.211 & \textbf{1.216} & 2.071 & \textbf{1.979} & \textbf{2.394} & 2.369 & \textbf{2.256} & \textbf{2.359} \\
    \bottomrule
    \end{tabular}%
  \label{tab:performance-aed}%
\end{table*}%

Network density, clustering coefficient, and the proportion of source nodes significantly impact model performance. Specifically, under comparable network density, a higher proportion of source nodes generally enhances model performance. For example, CollegeMsg and BA networks exhibit similar densities, but with source node proportions of 10\% and 5\%, respectively, resulting in F1 scores of 0.665 and 0.536 under the SI model. This can be attributed to the increased likelihood of correctly identifying source nodes with a higher proportion. On the other hand, when the source proportion is comparable, the connectivity characteristics of the network (e.g., density and clustering coefficient) also produce clear performance differences. For instance, the Jazz network (density 0.07) and the BA network (density 0.004) achieve F1 scores of 0.644 and 0.536, respectively, under the SI model with a 5\% source proportion. Likewise, for CollegeMsg and Cora-ML networks, which both use a 10\% source node proportion, the clustering coefficients are 0.087 and 0.240, with F1 scores of 0.665 and 0.654, respectively. This suggests that denser networks enhance node connectivity, providing richer topological information to the model and thereby improving predictive performance. The above findings are further validated on the Jazz datasets. Even with a lower source proportion, the model attains performance comparable to that on CollegeMsg, whose density and clustering coefficient are lower but whose source proportion is higher. This outcome further demonstrates that network density, clustering coefficient, and the source proportion jointly shape the model’s predictive behavior across different datasets.

Table \ref{tab:performance-real-world} illustrates the inference performance of our model in real-world information propagation scenarios. Due to its reliance on the global network structure, the IVGD and SDSA models face significant limitations in addressing this scenario. 
Apart from PDSL and BOSouL, other methods exhibit significantly degraded performance in scenarios with varying topological structures, approaching nearly complete collapse, which indicates substantial limitations of solely considering structural uncertainties in real-world scenes. In line with the findings on synthetic cascade datasets, PDSL consistently surpasses alternative approaches across all evaluation metrics except recall, regardless of whether the number of source nodes is fixed or unfixed. This demonstrates the superior generalization ability of PDSL in inferring propagation sources in real-world scenarios. Moreover, under the fixed-source setting, the reduced uncertainty in label distributions allows the model to exhibit stronger classification capability.

\subsection{Topology based Comparison}
To provide a comprehensive assessment of the localization performance from a topological perspective, we introduce the Average Error Distance (AED) as a structural evaluation metric, which calculates the average shortest path distance between each predicted source and its nearest ground-truth source in the graph. The definition is,
$$
AED = \min_{P \in permutation(\hat{S})} \frac{1}{K} \sum_{i=1}^{K} dist(s_i, p_i),
$$
where $K$ is the number of sources, $s_i$ is the ground truth source, $\hat{S}$ is the set of predicted sources and $P=\{p_1, p_2, \cdots, p_K\}$ is a permutation of $\hat{S}$. $dist(\cdot, \cdot)$ represents the shortest path.
Regarding the implementation of this metric, PDSL and several baselines are designed to infer sources without requiring prior knowledge of the exact number of source nodes. To ensure a fair and rigorous comparison, we select the Top-K nodes with the highest predicted probabilities as the inferred sources, where K is the actual number of ground-truth sources.

As illustrated in Table \ref{tab:performance-aed}, PDSL achieves the best performance in the majority of experimental scenarios. Although specific baseline methods, such as SDSA and BOSouL, demonstrate competitive results on certain datasets, their performance tends to exhibit noticeable fluctuations across varying network topologies or diffusion mechanisms. In contrast, the consistently stable performance of PDSL across diverse datasets suggests that integrating generative modeling with continuous-time dynamics provides the framework with robust generalization capabilities. This stability is particularly crucial for real-world deployment, where the underlying diffusion and network structures may be complex and heterogeneous.

\begin{figure*}
\centering
\includegraphics[width=0.95\textwidth]{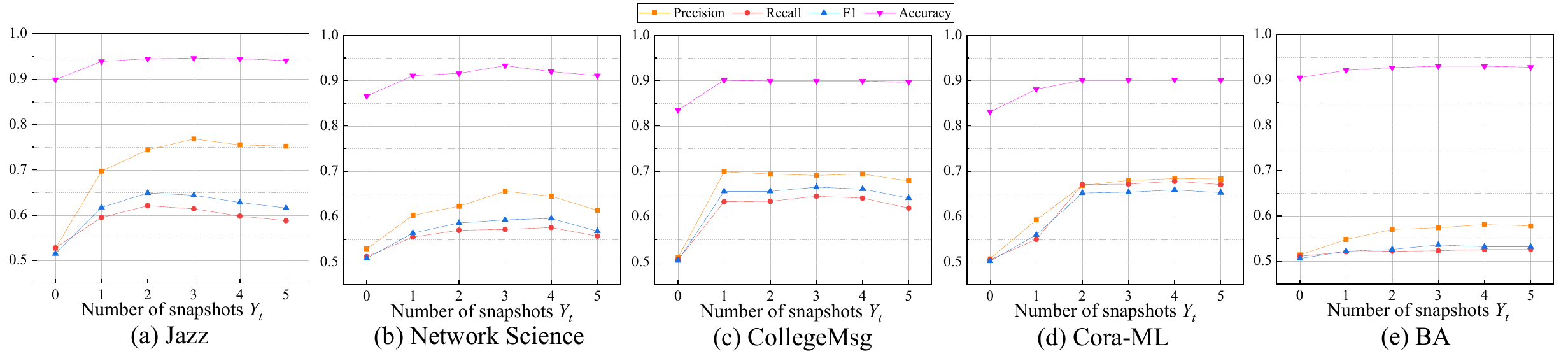}
\caption{The performance with different snapshot vectors under the SI diffusion mechanism.}
\label{fig:snapshot}
\end{figure*}

\subsection{Ablation Study}
We further perform ablation studies by comparing the PDSL with other variants to investigate the contribution of each sub-module.  propagation dynamics, data matching, and Graph Neural ODEs are three critical modules in our framework. For the propagation dynamics, we design two variants: 1) \textbf{w/o PD}, where propagation dynamics are completely removed, thus the source quantification module is implemented by the original VAEs. 2) \textbf{w/-Static}, where we restrict the input to solely the final propagation result $Y_T$, ignoring intermediate snapshots, to evaluate the model's capability under static observation. For the \textbf{w/o Gene}, we remove the generator module and directly utilize the latent representation $z$, produced by the encoder, as the candidate source. For the \textbf{w/o Matching}, we sample an initial diffusion source block $S_b$ from a Gaussian distribution. \textbf{w/o ODE} removes the Graph Neural ODEs module and only retains MLP in the forward propagation module. Finally, for the \textbf{w/o FP}, we remove the entire forward propagation module. The experimental results on 5 datasets are demonstrated in Table \ref{tab:ablation-si} and Table \ref{tab:ablation-glt}. 

Overall, the removal of any component results in performance degradation. This demonstrates that the cooperative interplay among these modules is pivotal to performance enhancement. The absence of propagation dynamics (w/o PD) results in a substantial decline in performance across all datasets. This phenomenon can be ascribed to the model's inability to adequately mitigate the uncertainties inherent in the propagation process. Mitigating such uncertainties is a primary focus of our model. 

Notably, even without access to intermediate snapshots, w/- Static achieves comparable or superior performance compared to state-of-the-art baselines. This implies that the effectiveness is not solely attributed to the availability of temporal data. Fundamentally, PDSL succeeds by learning the underlying diffusion mechanism to constrain source estimation, instead of simply mapping propagation results directly to source distributions. This capacity ensures that PDSL remains applicable to scenarios where only the final diffusion state or extremely sparse snapshots are available.

The substantial performance degradation observed in w/o Gene indicates that, although the latent space preserves essential information, it is still insufficient to directly characterize a valid source distribution. Therefore, the generator plays a crucial role in decoding latent features into candidate sources. In contrast, after removing the matching module (w/o Matching), the propagation sources are randomly initialized, yet the performance variation remains marginal. This result demonstrates the strong optimization capability of the proposed framework, which enables it to recover reliable source estimates even in the absence of explicit matching guidance.

Both the w/o ODE and w/o FP variants exhibit performance degradation across all datasets, which verifies the necessity of the forward propagation module. Notably, the w/o ODE variant even performs worse than the w/o FP variant, where the entire module is removed. This negative optimization phenomenon suggests that the complexity of propagation patterns requires the model to capture continuous evolutionary dynamics. The overly simplified architecture of the MLP is insufficient to model the multivariate continuous interactions involved in the diffusion process, thereby limiting the model's representation capacity. Conversely, ODE-based modeling can characterize the information evolution process more precisely, enhancing the model's generalization ability in complex scenarios.

\subsection{Sensitivity Analysis}
\begin{figure}[htbp]
\centering
\includegraphics[width=0.48\textwidth]{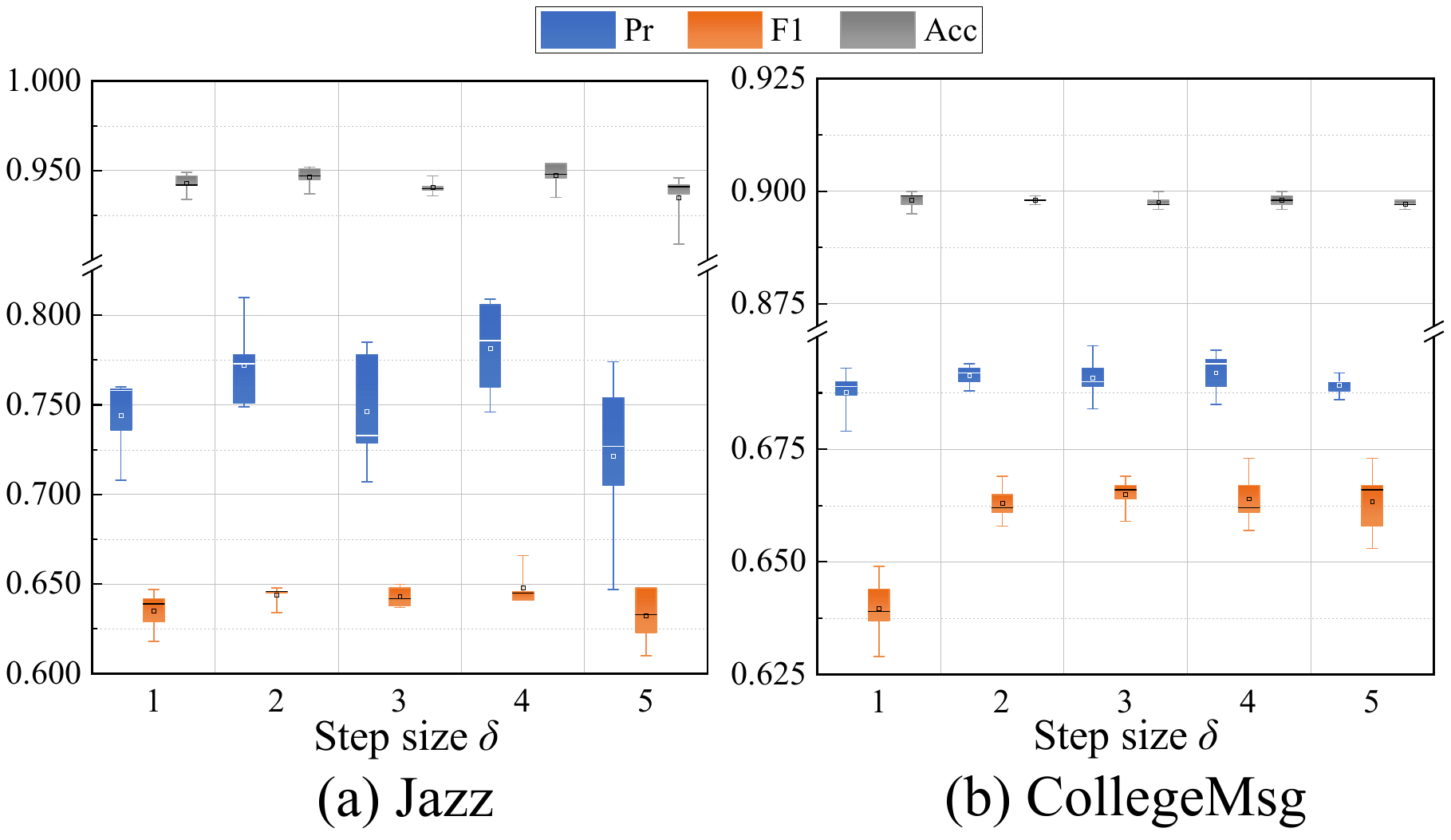}
\caption{The performance with different time step sizes under the SI diffusion mechanism.}
\label{fig:delta}
\end{figure}

Next, it is crucial to investigate the impact of hyperparameter configurations on model performance. In this part, we explore the number of snapshot vectors corresponding to the propagation dynamics and the step size corresponding to the forward propagation module. For snapshot vectors, we change the number from 0 to 5. For the step size, the numbers range from 1 to 5. The impacts on all scores are shown in Fig. \ref{fig:snapshot} and Fig. \ref{fig:delta}.

In the scenario where the count of snapshot vectors is zero, aligning with the w/o PD variant in the ablation study, the model exhibits notably weaker performance. The incorporation of a single snapshot vector (analogous to the w/-static variant in the ablation study), effectively conditioning the generation of latent diffusion sources solely on the final propagation result $Y_T$, leads to a clear performance improvement. This suggests that PDSL does not strictly rely on dense, continuous temporal dynamics. Instead, it can effectively capture the underlying diffusion mechanism even from very sparse observations. For the Jazz dataset, the F1 score increases substantially from 0.515 to 0.619, representing an improvement of approximately 20\%. A similar trend is observed on the Network Science dataset, where the F1 score rises from 0.508 to 0.564, corresponding to an improvement of around 10\%. It is worth noting that increasing the number of snapshot vectors does not always yield monotonic gains. Across most datasets, the best performance appears when using two to four snapshots, after which it stabilizes or shows a slight decline. This behavior may be attributed to the accumulation of redundant or noisy temporal information that reduces model stability. To balance predictive performance and computational efficiency, all experiments in this paper adopt a configuration of three snapshot vectors.

The time step size $\delta$ in the forward propagation module is a critical hyperparameter for modeling the dynamic evolution of nodes. When $\delta=1$, the model degenerates into a single-step propagation based on GNNs, implemented here using a graph convolutional network (GCN). As shown in Fig. \ref{fig:delta}, moderately increasing the time step size enhances model performance. Specifically, when $\delta$ increases from 1 to 2, the average Precision on the Jazz improves from 0.744 to 0.772 ($\approx3.6\%$), and the average F1 on the CollegeMsg rises from 0.639 to 0.663 ($\approx3.8\%$). These gains suggest that multi-step propagation better captures complex dynamic information. However, larger steps lead to performance degradation and significant fluctuations, likely due to accumulated noise during propagation, which undermines the model's generalization ability. Considering both effectiveness and efficiency, we set the time step size to $\delta=2$.

\begin{table*}[!htb]\scriptsize
    \caption{Ablation Study under SI Diffusion Pattern.}
    \centering
    \setlength{\tabcolsep}{0.8mm}{
    \begin{tabular}{c|c c c c|c c c c | c c c c | c c c c | c c c c}
    \toprule
   \multicolumn{1}{c|}{} & \multicolumn{4}{c|}{Jazz} &\multicolumn{4}{c|}{Network science} &\multicolumn{4}{c|}{CollegeMsg} &\multicolumn{4}{c|}{Cora-ML} &\multicolumn{4}{c}{BA} \\
    \midrule
    \multicolumn{1}{c|}{Method} & Pr & Re & F1 & Acc & Pr & Re & F1 & Acc & Pr & Re & F1 & Acc & Pr & Re & F1 & Acc & Pr & Re & F1 & Acc \\
    \midrule
    w/o PD & 0.527 & 0.528 & 0.515 & 0.899 & 0.529 & 0.512 & 0.508 & 0.866 & 0.510  & 0.505 & 0.503 & 0.835 & 0.507 & 0.505 & 0.502 & 0.831 & 0.514 & 0.511 & 0.506 & 0.905 \\
    w/-Static & 0.697 & 0.595 & 0.617 & 0.939 & 0.611 & 0.568 & 0.590  & 0.911 & 0.679 & 0.633 & 0.656 & 0.895 & 0.593 & 0.550  & 0.560  & 0.881 & 0.548 & 0.521 & 0.522 & 0.921 \\
    w/o Gene & 0.560 & 0.550 & 0.540 & 0.893 & 0.559 & 0.534 & 0.539 & 0.904 & 0.627 & 0.605 & 0.614 & 0.878 & 0.561 & 0.537 & 0.542 & 0.869 & 0.519 & 0.513 & 0.510 & 0.911 \\
    w/o Matching & 0.685 & 0.612 & 0.631 & 0.937 & 0.648 & 0.563 & 0.587 & 0.921 & 0.689 & 0.639 & 0.659 & 0.897 & 0.624 & 0.551 & 0.576 & 0.887 & 0.561 & 0.520  & 0.521 & 0.924 \\
    w/o ODE & 0.603 & 0.590  & 0.593 & 0.916 & 0.548 & 0.541 & 0.543 & 0.886 & 0.668 & 0.629 & 0.644 & 0.891 & 0.609 & \textbf{0.554} & 0.566 & 0.884 & 0.529 & 0.514 & 0.513 & 0.909 \\
    w/o FP & 0.713 & 0.592 & 0.617 & 0.927 & 0.629 & 0.552 & 0.578 & 0.914 & 0.682 & 0.633 & 0.649 & 0.885 & 0.615 & 0.534 & 0.568 & 0.884 & 0.558 & 0.511 & 0.523 & 0.921 \\
    Ours  & \textbf{0.768} & \textbf{0.614} & \textbf{0.644} & \textbf{0.946} & \textbf{0.656} & \textbf{0.572} & \textbf{0.593} & \textbf{0.933} & \textbf{0.691}  & \textbf{0.645} & \textbf{0.665} & \textbf{0.899} & \textbf{0.630}  & 0.553 & \textbf{0.584} & \textbf{0.891} & \textbf{0.574} & \textbf{0.523} & \textbf{0.536} & \textbf{0.930} \\
    \bottomrule
    \end{tabular}
    }
    \label{tab:ablation-si}
\end{table*}

\begin{table*}[!htb]\scriptsize
    \caption{Ablation Study under GLT Diffusion Pattern.}
    \centering
    \setlength{\tabcolsep}{0.8mm}{
    \begin{tabular}{c|c c c c|c c c c | c c c c | c c c c | c c c c}
    \toprule
   \multicolumn{1}{c|}{} & \multicolumn{4}{c|}{Jazz} &\multicolumn{4}{c|}{Network science} &\multicolumn{4}{c|}{CollegeMsg} &\multicolumn{4}{c|}{Cora-ML} &\multicolumn{4}{c}{BA} \\
    \midrule
    \multicolumn{1}{c|}{Method} & Pr & Re & F1 & Acc & Pr & Re & F1 & Acc & Pr & Re & F1 & Acc & Pr & Re & F1 & Acc & Pr & Re & F1 & Acc \\
    \midrule
    w/o PD & 0.568 & 0.534 & 0.532 & 0.905 & 0.523 & 0.509 & 0.504 & 0.887 & 0.508 & 0.506 & 0.504 & 0.831 & 0.514 & 0.508 & 0.506 & 0.840  & 0.516 & 0.512 & 0.507 & 0.905 \\
    w/-Static & 0.728 & 0.653 & 0.676 & 0.946 & 0.604 & 0.567 & 0.579 & 0.911 & 0.685 & 0.639 & 0.659 & 0.898 & 0.615 & 0.572 & 0.585 & 0.884 & 0.600 & 0.525 & 0.537 & 0.934 \\
    w/o Gene & 0.589 & 0.591 & 0.583 & 0.908 & 0.555 & 0.532 & 0.537 & 0.903 & 0.628 & 0.607 & 0.615 & 0.878 & 0.571 & 0.542 & 0.548 & 0.871 & 0.525 & 0.514 & 0.514 & 0.917 \\
    w/o Matching & 0.785 & 0.664 & 0.693 & 0.944 & 0.592 & 0.545 & 0.563 & 0.894 & 0.685 & 0.635 & 0.656 & 0.897 & 0.637 & 0.574 & 0.589 & 0.889 & 0.603 & 0.529 & 0.534 & 0.936 \\
    w/o ODE & 0.643 & 0.650  & 0.637 & 0.919 & 0.552 & 0.542 & 0.545 & 0.886 & 0.675 & 0.628 & 0.646 & 0.894 & 0.596 & 0.551 & 0.561 & 0.880  & 0.528 & 0.518 & 0.518 & 0.909 \\
    w/o FP & 0.704 & 0.668 & 0.662 & 0.921 & 0.641 & 0.566 & 0.585 & 0.921 & 0.678 & 0.632 & 0.651 & 0.887 & 0.629 & 0.564 & 0.581 & 0.879 & 0.611 & 0.523 & 0.529 & 0.927 \\
    Ours  & \textbf{0.803} & \textbf{0.681} & \textbf{0.719} & \textbf{0.957} & \textbf{0.667} & \textbf{0.569} & \textbf{0.591} & \textbf{0.936} & \textbf{0.689} & \textbf{0.642} & \textbf{0.661} & \textbf{0.898} & \textbf{0.645} & \textbf{0.575} & \textbf{0.593} & \textbf{0.897} & \textbf{0.636} & \textbf{0.531} & \textbf{0.542} & \textbf{0.941} \\
    \bottomrule
    \end{tabular}
    }
    \label{tab:ablation-glt}
\end{table*}

\begin{figure*}
\centering
\subfigure[LPSI]{\includegraphics[width=3cm]{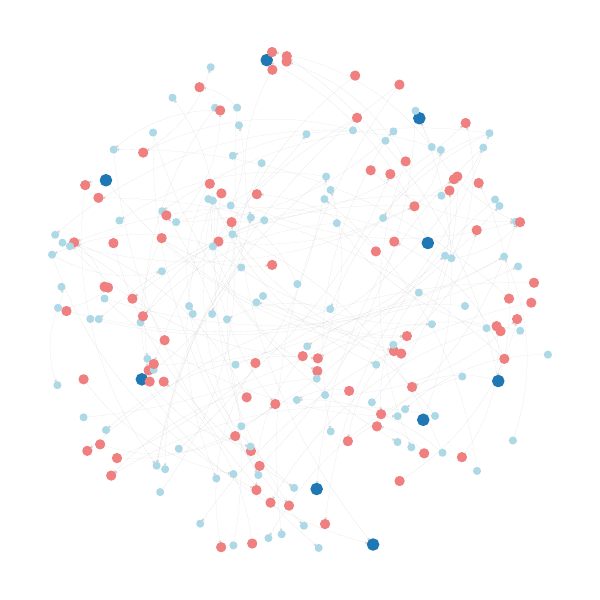}}
\subfigure[Netsleuth]{\includegraphics[width=3cm]{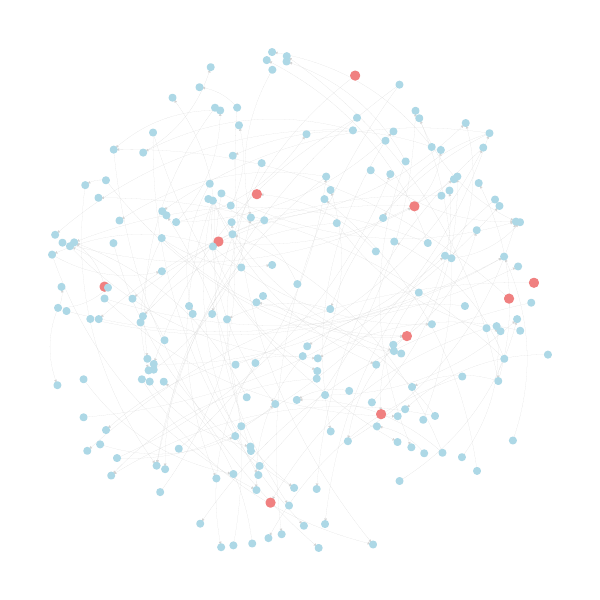}}
\subfigure[OJC]{\includegraphics[width=3cm]{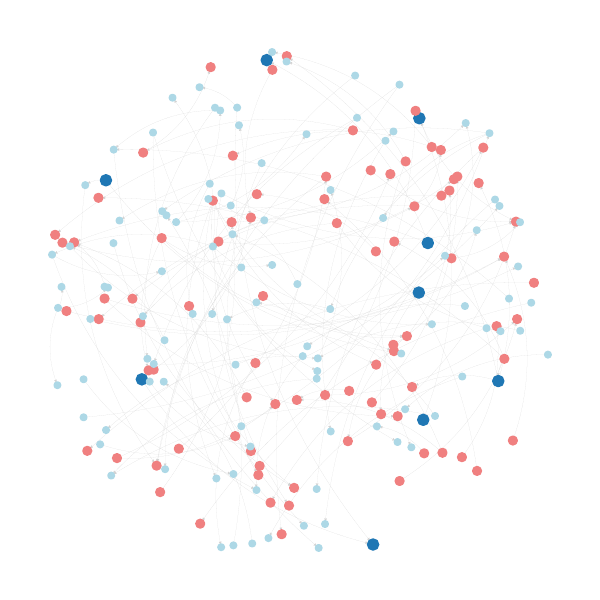}}
\subfigure[GCNSI]{\includegraphics[width=3cm]{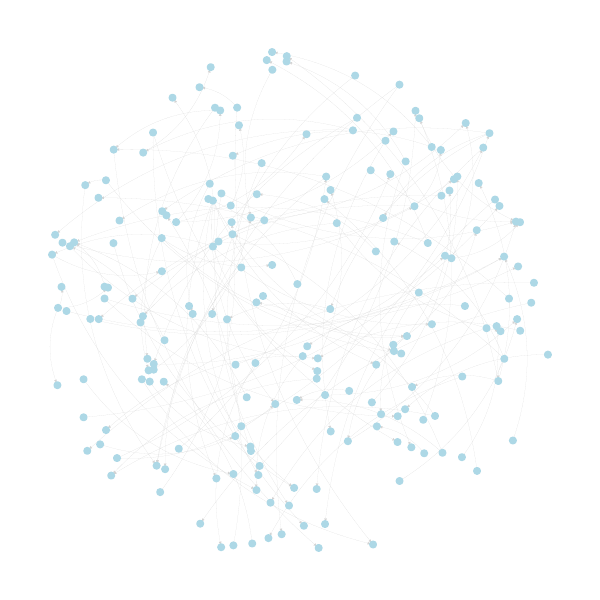}}
\subfigure[IVGD]{\includegraphics[width=3cm]{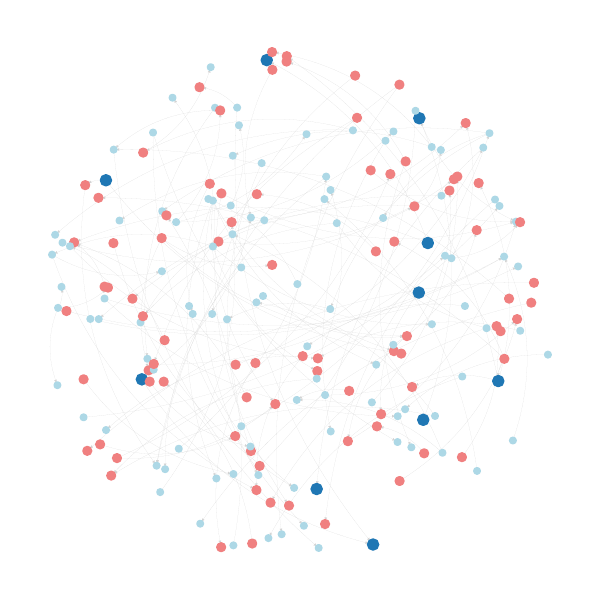}}
\subfigure[SLVAE]{\includegraphics[width=3cm]{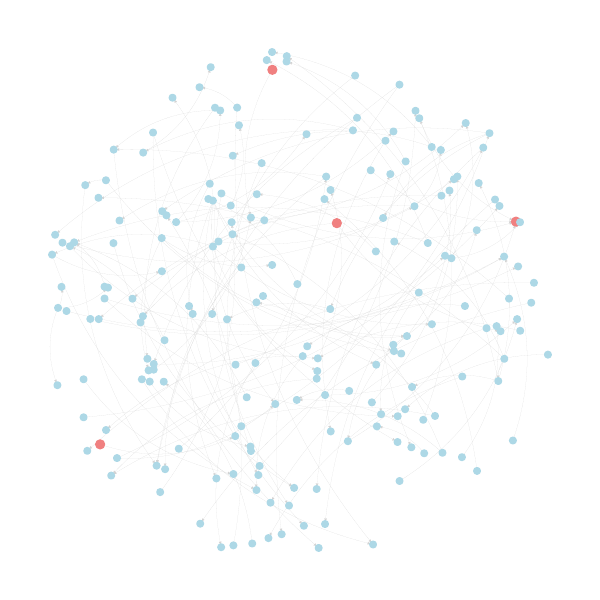}}
\subfigure[BOSouL]{\includegraphics[width=3cm]{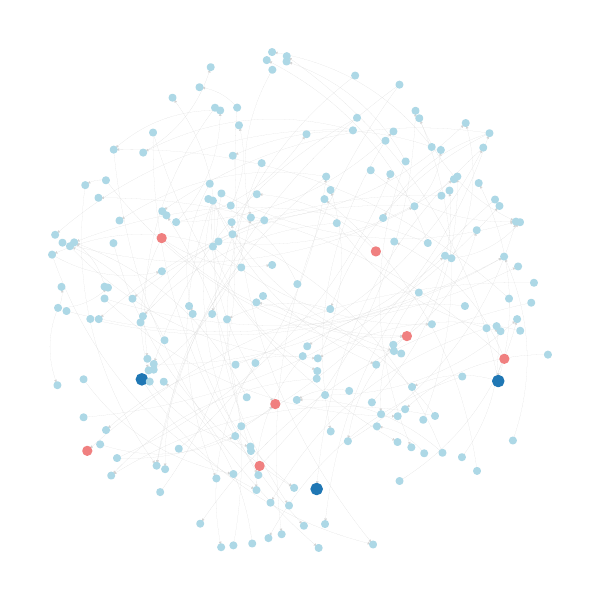}}
\subfigure[SDSA]{\includegraphics[width=3cm]{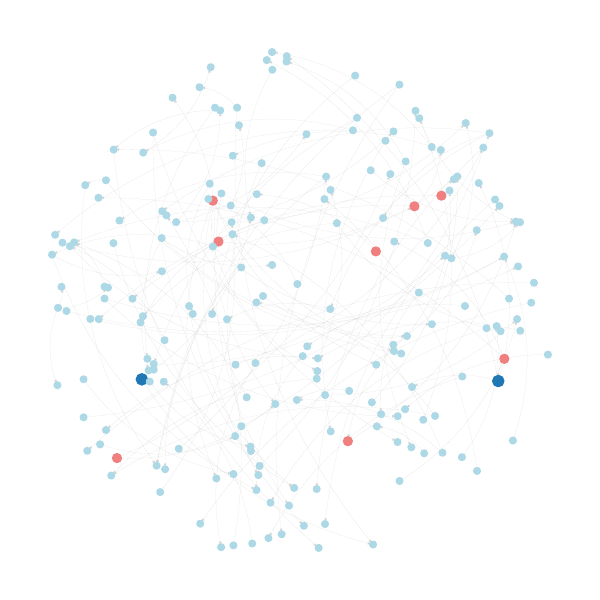}}
\subfigure[PDSL]{\includegraphics[width=3cm]{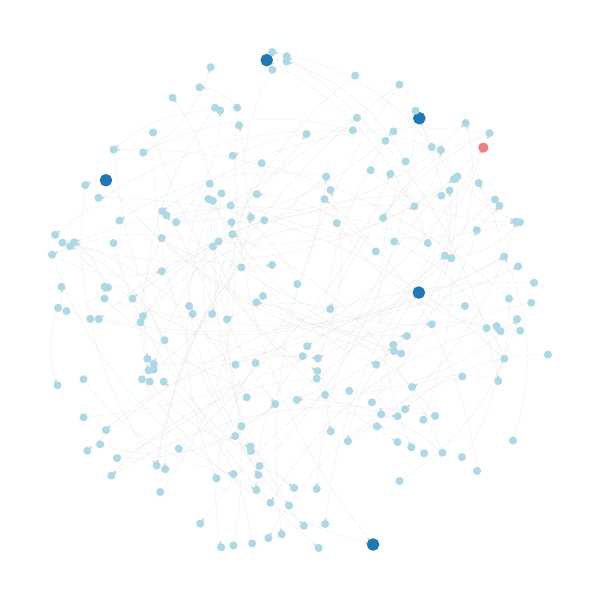}}
\subfigure[True]{\includegraphics[width=3cm]{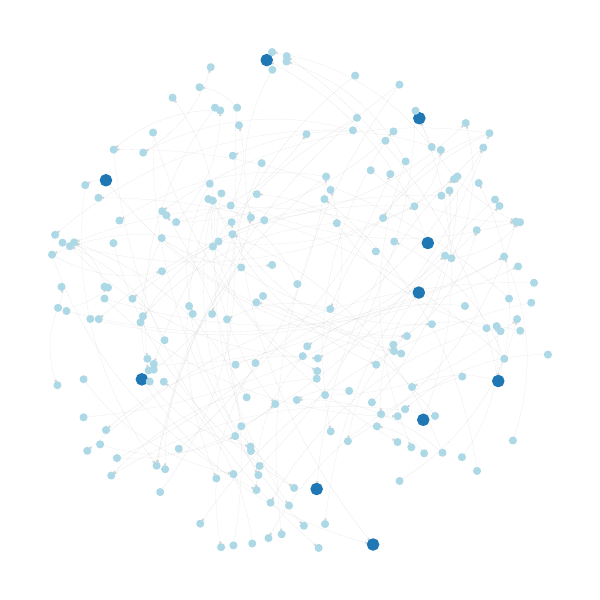}}
\caption{Visualizations of Jazz dataset for all methods and the ground truth. The correctly predicted sources are marked in blue, while the incorrectly predicted sources are marked in light red. The remaining nodes are marked in light blue.}
\label{fig:visualization}
\end{figure*}

\subsection{Complexity Analysis}
Our framework consists of two primary components: Source Quantification and Forward Propagation. The first component comprises MLP based encoder and generator. The encoder maps an input of dimension $|V|$ to a latent space characterized by $\mu$ and $\sigma$, with a total dimension of $2K$, incurring a complexity of $\mathcal{O}((T-1)|V|H + H^2 + 2HK)$, where $H$ denotes the hidden layer dimension. The generator transforms the latent variable $z$ into the potential source vector $s^*$, with a complexity of $\mathcal{O}((K+(T-1)|V|)H + H^2 + H|V|)$, dominated by $\mathcal{O}((T-1)|V|H)$. The second component employs a Graph Neural ODEs. Each GNN layer, processing $d$-dimensional features, involves feature transformation with complexity $\mathcal{O}(|V|d^2)$ and neighborhood aggregation with complexity $\mathcal{O}(|E|d)$, resulting in a total complexity for $L$ GNN layers of $\mathcal{O}(L(|V|d^2 + |E|d))$. The ODE solver discretizes the continuous process into $\delta$ steps, yielding a complexity for this module of $\mathcal{O}(\delta L(|V|d^2 + |E|d))$.

In the training phase, the overall complexity is dominated by the Graph Neural ODEs module, which is responsible for modeling continuous propagation. The complexity is $\mathcal{O}(\delta L(|V|d^2 + |E|d))$, which is linear with respect to both the number of nodes $|V|$ and the number of edges $|E|$. In sparse real-world networks where $|E| << |V|^2$, this ensures that the training cost remains manageable even as the network size increases. The inference phase employs a matching strategy followed by an $N$-step iterative optimization. With complexities of $\mathcal{O}(B|V|)$ and $\mathcal{O}(N \delta L |V| d^2)$ respectively, this yields a total computational cost of $\mathcal{O}(B|V| + N \delta L |V| d^2)$ , where $B$ is the batch size. Similar to the training phase, the inference cost scales linearly with the number of nodes $|V|$. This linear dependency guarantees that PDSL can handle large-scale source localization tasks.

\begin{figure}
    \centering
    \includegraphics[width=0.8\linewidth]{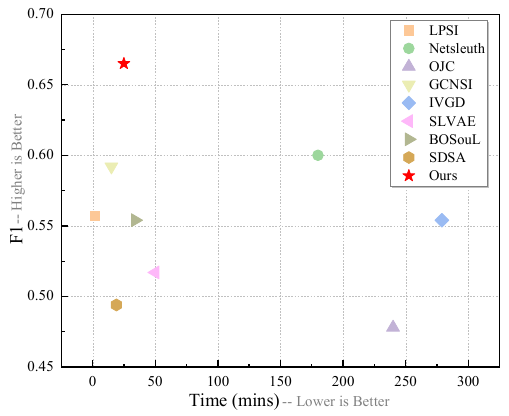}
    \caption{Comparison of runtime and F1-score performance on CollegeMsg (SI Model), where the x-axis represents the running time and the y-axis denotes the F1-score.}
    \label{fig:time}
\end{figure}

We further verify the efficiency of our proposed method by conducting a detailed runtime analysis on the CollegeMsg dataset. Since traditional heuristic baselines are typically CPU-bound, we report the results in the same CPU environment for a fair comparison. The results are plotted in Figure \ref{fig:time}, observations indicate that our model achieves a substantial performance gain with only a negligible increase in runtime. Given that source localization is typically a post-hoc analysis task where precision is of primary importance, our method achieves a well-balanced trade-off between efficiency and effectiveness.

\subsection{Case Study}
Finally, Fig. \ref{fig:visualization} presents a comprehensive case study of the reconstruction performance of diffusion sources under SI diffusion pattern across various comparative methods. For enhanced clarity, the Jazz dataset with a relatively small graph size, has been selected for this illustration. Specifically, our proposed PDSL demonstrates a distribution closely resembling the real dataset. LPSI, OJC, and IVGD exhibit a propensity to overestimate the number of sources, thereby achieving higher Recall metrics. GCNSI omits a significant portion of the sources due to its inadequate management of label imbalance. Meanwhile, SLVAE's oversight in addressing uncertainty leads to inaccuracies in source prediction. Despite Netsleuth and BOSouL preset the source counts, the uncertainty in the propagation process results in suboptimal prediction outcomes. In summary, these observations align with the quantitative results in Table \ref{tab:performance-si}, Table \ref{tab:performance-glt}, and Table \ref{tab:performance-sir}.

\section{Conclusion and Discussion}\label{sec:conclusion}
Source localization is of paramount importance for uncovering information propagation within networks. In this paper, we propose a novel framework, PDSL, which first introduces the integration of a deep generative model with propagation dynamics to reduce the inherent uncertainty in information dissemination processes. PDSL is anchored in a generative model that effectively approximates the distribution of sources, utilizing propagation dynamics to mitigate uncertainty induced by the stochastic nature of diffusion processes. In addition, a Graph Neural ODEs based forward propagation model is designed to simulate information dissemination without relying on predefined diffusion mechanisms. The framework benefits from a novel approach to learning and inferring source distribution, surpassing traditional methods. The efficiency of the proposed model is validated through extensive experiments and studies on real-world datasets, consistently demonstrating superior performance in diverse application scenarios. 

As a data-driven framework, the performance of PDSL is inherently linked to the quality and representativeness of the training cascades. In scenarios with highly sparse data or uncaptured stochasticity, inferential accuracy may be constrained. However, PDSL incorporates several design elements to mitigate this dependency. Specifically, the integration of Graph Neural ODEs (GODE) allows for continuous-time modeling, providing a strong inductive bias to infer underlying dynamics from noisy or irregularly sampled data. Furthermore, the generative nature of our framework explicitly models propagation stochasticity, enhancing robustness against non-representative outliers. While extreme data scarcity remains a challenge, PDSL effectively balances data-driven learning with theoretical robustness. Future research is likely to explore deeper integration with reinforcement learning and causal inference to further enhance source localization capabilities.

\appendices
\section{Proof of the Equation (\ref{equ:L_KL})}
\label{appendix}
In the following, we present the detailed derivation of Equation (\ref{equ:L_KL}).
\begin{equation}\small
\nonumber
\begin{aligned}
     & KL \Big( q_\theta(z|s, Y_t, G) || p(z|Y_t, G)  \Big) \\
     =& KL \Big( N\big(\mu_\theta(s, Y_t, G), \sigma_\theta^2(s, Y_t, G)\big) || N(0,I)  \Big) \\
     =& \int \frac{1}{\sqrt{2\pi\sigma_\theta^2}}\text{exp} \big(-\frac{(z-\mu_\theta)^2}{2\sigma_\theta^2} \big) \ln \frac{\frac{1}{\sqrt{2\pi\sigma_\theta^2}} \text{exp}(-\frac{(z-\mu_\theta)^2}{2\sigma_\theta^2})}{\frac{1}{\sqrt{2\pi}} \text{exp}(-\frac{z^2}{2})} dz \\
     =& \int \frac{1}{\sqrt{2\pi\sigma_\theta^2}}\text{exp} \big(-\frac{(z-\mu_\theta)^2}{2\sigma_\theta^2} \big) \ln \big( \frac{1}{\sqrt{\sigma_\theta^2}} \text{exp}(\frac{z^2}{2}-\frac{(z-\mu_\theta)^2}{2\sigma_\theta^2}) \big) dz \\
     =& -\frac{1}{2} \ln \sigma_\theta^2 \int p(z) dz + \frac{1}{2} z^2 \int p(z) dz - \frac{1}{2\sigma_\theta^2} \int (z - \mu_\theta)^2 p(z) dz,
\end{aligned}
\end{equation}
where $p(z) = \frac{1}{\sqrt{2\pi\sigma_\theta^2}}\text{exp}(-\frac{(z-\mu_\theta)^2}{2\sigma_\theta^2})$ is the probability density function. The first term $\int p(z) dz =1$. The second term corresponds to computing the second moment $\mathbb{E}_p[z^2]$, where for $z \sim N(\mu_\theta, \sigma_\theta^2)$, we have $\mathbb{E}[z^2] = Var(z) + (\mathbb{E}[z])^2 = \sigma_\theta^2 + \mu_\theta^2$. The third term corresponds to computing the variance $\mathbb{E}_p[(z - \mu_\theta)^2] = \sigma_\theta^2$. Thus the result is,
\begin{equation}
\nonumber
    KL \Big( q_\theta(z|s, Y_t, G) || p(z|Y_t, G)  \Big) = \frac{1}{2}(-\ln \sigma_{\theta}^2 + \mu_{\theta}^2 + \sigma_{\theta}^2 -1).
\end{equation}



\ifCLASSOPTIONcaptionsoff
  \newpage
\fi

\bibliographystyle{IEEEtran}
\bibliography{software}

@inproceedings{ru2023inferring,
  title={Inferring patient zero on temporal networks via graph neural networks},
  author={Ru, Xiaolei and Moore, Jack Murdoch and Zhang, Xin-Ya and Zeng, Yeting and Yan, Gang},
  booktitle={Proceedings of the AAAI Conference on Artificial Intelligence},
  volume={37},
  number={8},
  pages={9632--9640},
  year={2023}
}

@article{li2012multiple,
  title={Multiple Location Profiling for Users and Relationships from Social Network and Content},
  author={Li, Rui and Wang, Shengjie and Chen-Chuan, Kevin},
  journal={Proceedings of the VLDB Endowment},
  volume={5},
  number={11},
  year={2012}
}

@article{shah2011rumors,
  title={Rumors in a network: Who's the culprit?},
  author={Shah, Devavrat and Zaman, Tauhid},
  journal={IEEE Transactions on information theory},
  volume={57},
  number={8},
  pages={5163--5181},
  year={2011},
  publisher={IEEE}
}

@article{jiang2015k,
  title={K-center: An approach on the multi-source identification of information diffusion},
  author={Jiang, Jiaojiao and Wen, Sheng and Yu, Shui and Xiang, Yang and Zhou, Wanlei},
  journal={IEEE Transactions on Information Forensics and Security},
  volume={10},
  number={12},
  pages={2616--2626},
  year={2015},
  publisher={IEEE}
}

@inproceedings{zhu2016information,
  title={Information source detection in networks: Possibility and impossibility results},
  author={Zhu, Kai and Ying, Lei},
  booktitle={IEEE INFOCOM 2016-The 35th Annual IEEE International Conference on Computer Communications},
  pages={1--9},
  year={2016},
  organization={IEEE}
}

@article{luo2014identify,
  title={How to identify an infection source with limited observations},
  author={Luo, Wuqiong and Tay, Wee Peng and Leng, Mei},
  journal={IEEE Journal of Selected Topics in Signal Processing},
  volume={8},
  number={4},
  pages={586--597},
  year={2014},
  publisher={IEEE}
}

@article{hodas2014simple,
  title={The simple rules of social contagion},
  author={Hodas, Nathan O and Lerman, Kristina},
  journal={Scientific reports},
  volume={4},
  number={1},
  pages={4343},
  year={2014},
  publisher={Nature Publishing Group UK London}
}

@inproceedings{hegde2018algorithmic,
  title={Algorithmic aspects of inverse problems using generative models},
  author={Hegde, Chinmay},
  booktitle={2018 56th Annual Allerton Conference on Communication, Control, and Computing (Allerton)},
  pages={166--172},
  year={2018},
  organization={IEEE}
}

@inproceedings{peng2021generating,
  title={Generating diverse structure for image inpainting with hierarchical VQ-VAE},
  author={Peng, Jialun and Liu, Dong and Xu, Songcen and Li, Houqiang},
  booktitle={Proceedings of the IEEE/CVF conference on computer vision and pattern recognition},
  pages={10775--10784},
  year={2021}
}

@article{duan2023qarv,
  title={Qarv: Quantization-aware resnet vae for lossy image compression},
  author={Duan, Zhihao and Lu, Ming and Ma, Jack and Huang, Yuning and Ma, Zhan and Zhu, Fengqing},
  journal={IEEE Transactions on Pattern Analysis and Machine Intelligence},
  year={2023},
  publisher={IEEE}
}

@article{sohn2015learning,
  title={Learning structured output representation using deep conditional generative models},
  author={Sohn, Kihyuk and Lee, Honglak and Yan, Xinchen},
  journal={Advances in neural information processing systems},
  volume={28},
  year={2015}
}

@article{kingma2013auto,
  title={Auto-encoding variational bayes},
  author={Kingma, Diederik P and Welling, Max},
  journal={arXiv preprint arXiv:1312.6114},
  year={2013}
}

@inproceedings{prakash2012spotting,
  title={Spotting culprits in epidemics: How many and which ones?},
  author={Prakash, B Aditya and Vreeken, Jilles and Faloutsos, Christos},
  booktitle={2012 IEEE 12th international conference on data mining},
  pages={11--20},
  year={2012},
  organization={IEEE}
}

@inproceedings{wang2017multiple,
  title={Multiple source detection without knowing the underlying propagation model},
  author={Wang, Zheng and Wang, Chaokun and Pei, Jisheng and Ye, Xiaojun},
  booktitle={Proceedings of the AAAI Conference on Artificial Intelligence},
  volume={31},
  number={1},
  year={2017}
}

@inproceedings{zhu2017catch,
  title={Catch’em all: Locating multiple diffusion sources in networks with partial observations},
  author={Zhu, Kai and Chen, Zhen and Ying, Lei},
  booktitle={Proceedings of the AAAI Conference on Artificial Intelligence},
  volume={31},
  number={1},
  year={2017}
}

@inproceedings{dong2019multiple,
  title={Multiple rumor source detection with graph convolutional networks},
  author={Dong, Ming and Zheng, Bolong and Quoc Viet Hung, Nguyen and Su, Han and Li, Guohui},
  booktitle={Proceedings of the 28th ACM international conference on information and knowledge management},
  pages={569--578},
  year={2019}
}

@inproceedings{wang2022invertible,
  title={An invertible graph diffusion neural network for source localization},
  author={Wang, Junxiang and Jiang, Junji and Zhao, Liang},
  booktitle={Proceedings of the ACM Web Conference 2022},
  pages={1058--1069},
  year={2022}
}

@inproceedings{ling2022source,
  title={Source localization of graph diffusion via variational autoencoders for graph inverse problems},
  author={Ling, Chen and Jiang, Junji and Wang, Junxiang and Liang, Zhao},
  booktitle={Proceedings of the 28th ACM SIGKDD conference on knowledge discovery and data mining},
  pages={1010--1020},
  year={2022}
}

@inproceedings{zhang2024multiple,
  title={Multiple-Source Localization from a Single-Snapshot Observation Using Graph Bayesian Optimization},
  author={Zhang, Zonghan and Zhang, Zijian and Chen, Zhiqian},
  booktitle={Proceedings of the AAAI Conference on Artificial Intelligence},
  volume={38},
  number={20},
  pages={22538--22546},
  year={2024}
}

@inproceedings{zang2015topic,
  title={Topic-aware source locating in social networks},
  author={Zang, Wenyu and Zhou, Chuan and Guo, Li and Zhang, Peng},
  booktitle={Proceedings of the 24th International Conference on World Wide Web},
  pages={141--142},
  year={2015}
}

@article{shah2020finding,
  title={Finding patient zero: Learning contagion source with graph neural networks},
  author={Shah, Chintan and Dehmamy, Nima and Perra, Nicola and Chinazzi, Matteo and Barab{\'a}si, Albert-L{\'a}szl{\'o} and Vespignani, Alessandro and Yu, Rose},
  journal={arXiv preprint arXiv:2006.11913},
  year={2020}
}

@article{shu2021information,
  title={Information source estimation with multi-channel graph neural network},
  author={Shu, Xincheng and Yu, Bin and Ruan, Zhongyuan and Zhang, Qingpeng and Xuan, Qi},
  journal={Graph Data Mining: Algorithm, Security and Application},
  pages={1--27},
  year={2021},
  publisher={Springer}
}

@article{xu2024pgsl,
  title={PGSL: A probabilistic graph diffusion model for source localization},
  author={Xu, Xovee and Qian, Tangjiang and Xiao, Zhe and Zhang, Ni and Wu, Jin and Zhou, Fan},
  journal={Expert Systems with Applications},
  volume={238},
  pages={122028},
  year={2024},
  publisher={Elsevier}
}

@article{spinelli2017general,
  title={A general framework for sensor placement in source localization},
  author={Spinelli, Brunella and Celis, L Elisa and Thiran, Patrick},
  journal={IEEE Transactions on Network Science and Engineering},
  volume={6},
  number={2},
  pages={86--102},
  year={2017},
  publisher={IEEE}
}

@article{paluch2020optimizing,
  title={Optimizing sensors placement in complex networks for localization of hidden signal source: A review},
  author={Paluch, Robert and Gajewski, {\L}ukasz G and Ho{\l}yst, Janusz A and Szymanski, Boleslaw K},
  journal={Future Generation Computer Systems},
  volume={112},
  pages={1070--1092},
  year={2020},
  publisher={Elsevier}
}

@inproceedings{wang2022rapid,
  title={A rapid source localization method in the early stage of large-scale network propagation},
  author={Wang, Zhen and Hou, Dongpeng and Gao, Chao and Huang, Jiajin and Xuan, Qi},
  booktitle={Proceedings of the ACM web conference 2022},
  pages={1372--1380},
  year={2022}
}

@inproceedings{wang2023lightweight,
  title={Lightweight source localization for large-scale social networks},
  author={Wang, Zhen and Hou, Dongpeng and Gao, Chao and Li, Xiaoyu and Li, Xuelong},
  booktitle={Proceedings of the ACM Web Conference 2023},
  pages={286--294},
  year={2023}
}

@article{chai2021information,
  title={Information sources estimation in time-varying networks},
  author={Chai, Yun and Wang, Youguo and Zhu, Liang},
  journal={IEEE Transactions on Information Forensics and Security},
  volume={16},
  pages={2621--2636},
  year={2021},
  publisher={IEEE}
}

@inproceedings{dawkins2021diffusion,
  title={Diffusion source identification on networks with statistical confidence},
  author={Dawkins, Quinlan E and Li, Tianxi and Xu, Haifeng},
  booktitle={International Conference on Machine Learning},
  pages={2500--2509},
  year={2021},
  organization={PMLR}
}

@article{zhu2022locating,
  title={Locating multi-sources in social networks with a low infection rate},
  author={Zhu, Peican and Cheng, Le and Gao, Chao and Wang, Zhen and Li, Xuelong},
  journal={IEEE Transactions on Network Science and Engineering},
  volume={9},
  number={3},
  pages={1853--1865},
  year={2022},
  publisher={IEEE}
}

@article{goldenberg2001talk,
  title={Talk of the network: A complex systems look at the underlying process of word-of-mouth},
  author={Goldenberg, Jacob and Libai, Barak and Muller, Eitan},
  journal={Marketing letters},
  volume={12},
  pages={211--223},
  year={2001},
  publisher={Springer}
}

@article{chang2018maximum,
  title={Maximum a posteriori estimation for information source detection},
  author={Chang, Biao and Chen, Enhong and Zhu, Feida and Liu, Qi and Xu, Tong and Wang, Zhefeng},
  journal={IEEE Transactions on Systems, Man, and Cybernetics: Systems},
  volume={50},
  number={6},
  pages={2242--2256},
  year={2018},
  publisher={IEEE}
}

@article{yan2024diffusion,
  title={Diffusion model for graph inverse problems: Towards effective source localization on complex networks},
  author={Yan, Xin and Fang, Hui and He, Qiang},
  journal={Advances in Neural Information Processing Systems},
  volume={36},
  year={2024}
}

@article{lazer2018science,
  title={The science of fake news},
  author={Lazer, David MJ and Baum, Matthew A and Benkler, Yochai and Berinsky, Adam J and Greenhill, Kelly M and Menczer, Filippo and Metzger, Miriam J and Nyhan, Brendan and Pennycook, Gordon and Rothschild, David and others},
  journal={Science},
  volume={359},
  number={6380},
  pages={1094--1096},
  year={2018},
  publisher={American Association for the Advancement of Science}
}

@article{cai2023adam,
  title={ADAM: an adaptive DDoS attack mitigation scheme in software-defined cyber-physical system},
  author={Cai, Tianyang and Jia, Tao and Adepu, Sridhar and Li, Yuqi and Yang, Zheng},
  journal={IEEE Transactions on Industrial Informatics},
  volume={19},
  number={6},
  pages={7802--7813},
  year={2023},
  publisher={IEEE}
}

@inproceedings{luo2013estimating,
  title={Estimating infection sources in a network with incomplete observations},
  author={Luo, Wuqiong and Tay, Wee Peng},
  booktitle={2013 IEEE Global Conference on Signal and Information Processing},
  pages={301--304},
  year={2013},
  organization={IEEE}
}

@article{cencetti2023distinguishing,
  title={Distinguishing simple and complex contagion processes on networks},
  author={Cencetti, Giulia and Contreras, Diego Andr{\'e}s and Mancastroppa, Marco and Barrat, Alain},
  journal={Physical Review Letters},
  volume={130},
  number={24},
  pages={247401},
  year={2023},
  publisher={APS}
}

@article{vosoughi2018spread,
  title={The spread of true and false news online},
  author={Vosoughi, Soroush and Roy, Deb and Aral, Sinan},
  journal={science},
  volume={359},
  number={6380},
  pages={1146--1151},
  year={2018},
  publisher={American Association for the Advancement of Science}
}

@article{epstein2023social,
  title={The social media context interferes with truth discernment},
  author={Epstein, Ziv and Sirlin, Nathaniel and Arechar, Antonio and Pennycook, Gordon and Rand, David},
  journal={Science Advances},
  volume={9},
  number={9},
  pages={eabo6169},
  year={2023},
  publisher={American Association for the Advancement of Science}
}

@article{chen2023modeling,
  title={Modeling and analyzing malware propagation over wireless networks based on hypergraphs},
  author={Chen, Jiaxing and Sun, Shiwen and Xia, Chengyi and Shi, Dinghua and Chen, Guanrong},
  journal={IEEE Transactions on Network Science and Engineering},
  volume={10},
  number={6},
  pages={3767--3778},
  year={2023},
  publisher={IEEE}
}

@article{duan2025modeling,
  title={Modeling and Predicting Malware Propagation in Double-Layer Computer Networks},
  author={Duan, Dongli and Yang, Xiaohao and Lv, Changchun and Cai, Zhiqiang},
  journal={IEEE Transactions on Network Science and Engineering},
  year={2025},
  publisher={IEEE}
}

@article{chen2018neural,
  title={Neural ordinary differential equations},
  author={Chen, Ricky TQ and Rubanova, Yulia and Bettencourt, Jesse and Duvenaud, David K},
  journal={Advances in neural information processing systems},
  volume={31},
  year={2018}
}

@article{poli2019graph,
  title={Graph neural ordinary differential equations},
  author={Poli, Michael and Massaroli, Stefano and Park, Junyoung and Yamashita, Atsushi and Asama, Hajime and Park, Jinkyoo},
  journal={arXiv preprint arXiv:1911.07532},
  year={2019}
}

@article{allen1994some,
  title={Some discrete-time SI, SIR, and SIS epidemic models},
  author={Allen, Linda JS},
  journal={Mathematical biosciences},
  volume={124},
  number={1},
  pages={83--105},
  year={1994},
  publisher={Elsevier}
}

@article{ran2020generalized,
  title={A generalized linear threshold model for an improved description of the spreading dynamics},
  author={Ran, Yijun and Deng, Xiaomin and Wang, Xiaomeng and Jia, Tao},
  journal={Chaos: An Interdisciplinary Journal of Nonlinear Science},
  volume={30},
  number={8},
  year={2020},
  publisher={AIP Publishing}
}

@inproceedings{xhonneux2020continuous,
  title={Continuous graph neural networks},
  author={Xhonneux, Louis-Pascal and Qu, Meng and Tang, Jian},
  booktitle={International conference on machine learning},
  pages={10432--10441},
  year={2020},
  organization={PMLR}
}

@inproceedings{chamberlain2021grand,
  title={Grand: Graph neural diffusion},
  author={Chamberlain, Ben and Rowbottom, James and Gorinova, Maria I and Bronstein, Michael and Webb, Stefan and Rossi, Emanuele},
  booktitle={International conference on machine learning},
  pages={1407--1418},
  year={2021},
  organization={PMLR}
}

@article{zhao2024mase,
  title={MASE: Multi-attribute source estimator for epidemic transmission in complex networks},
  author={Zhao, Jie and Cheong, Kang Hao},
  journal={IEEE Transactions on Systems, Man, and Cybernetics: Systems},
  volume={54},
  number={6},
  pages={3308--3320},
  year={2024},
  publisher={IEEE}
}

@article{hou2024random,
  title={Random full-order-coverage based rapid source localization with limited observations for large-scale networks},
  author={Hou, Dongpeng and Gao, Chao and Wang, Zhen and Li, Xiaoyu and Li, Xuelong},
  journal={IEEE Transactions on Network Science and Engineering},
  volume={11},
  number={5},
  pages={4213--4226},
  year={2024},
  publisher={IEEE}
}

@article{zhao2025enhanced,
  title={Enhanced epidemic control: Community-based observer placement and source tracing},
  author={Zhao, Jie and Cheong, Kang Hao},
  journal={IEEE Transactions on Systems, Man, and Cybernetics: Systems},
  year={2025},
  publisher={IEEE}
}

@article{hou2025fgssi,
  title={Fgssi: a feature-enhanced framework with transferability for sequential source identification},
  author={Hou, Dongpeng and Gao, Chao and Wang, Zhen and Li, Xuelong},
  journal={IEEE Transactions on Dependable and Secure Computing},
  year={2025},
  publisher={IEEE}
}

@inproceedings{hou2024new,
  title={New Localization Frameworks: User-centric Approaches to Source Localization in Real-world Propagation Scenarios},
  author={Hou, Dongpeng and Wang, Yuchen and Gao, Chao and Li, Xianghua and Wang, Zhen},
  booktitle={Proceedings of the 33rd ACM International Conference on Information and Knowledge Management},
  pages={839--848},
  year={2024}
}

@article{runge1895numerische,
  title={{\"U}ber die numerische Aufl{\"o}sung von Differentialgleichungen},
  author={Runge, Carl},
  journal={Mathematische Annalen},
  volume={46},
  number={2},
  pages={167--178},
  year={1895},
  publisher={Springer}
}

@article{liu2023diffusion,
  title={Diffusion source inference for large-scale complex networks based on network percolation},
  author={Liu, Yang and Wang, Xiaoqi and Wang, Xi and Wang, Zhen and Kurths, J{\"u}rgen},
  journal={IEEE Transactions on Neural Networks and Learning Systems},
  year={2023},
  publisher={IEEE}
}

@article{wei2022modeling,
  title={Modeling the uncertainty of information propagation for rumor detection: A neuro-fuzzy approach},
  author={Wei, Lingwei and Hu, Dou and Zhou, Wei and Wang, Xin and Hu, Songlin},
  journal={IEEE transactions on neural networks and learning systems},
  volume={35},
  number={2},
  pages={2522--2533},
  year={2022},
  publisher={IEEE}
}

@article{wang2024modality,
  title={Modality perception learning-based determinative factor discovery for multimodal fake news detection},
  author={Wang, Boyue and Wu, Guangchao and Li, Xiaoyan and Gao, Junbin and Hu, Yongli and Yin, Baocai},
  journal={IEEE Transactions on Neural Networks and Learning Systems},
  year={2024},
  publisher={IEEE}
}

@article{yang2022learning,
  title={Learning deep generative clustering via mutual information maximization},
  author={Yang, Xiaojiang and Yan, Junchi and Cheng, Yu and Zhang, Yizhe},
  journal={IEEE Transactions on Neural Networks and Learning Systems},
  volume={34},
  number={9},
  pages={6263--6275},
  year={2022},
  publisher={IEEE}
}

@article{saha2025gen,
  title={Gen-GraphEx: Generative In-Distribution Graph Explanations for Time-Efficient Model-Level Interpretability of GNNs},
  author={Saha, Sayan and Das, Monidipa and Bandyopadhyay, Sanghamitra},
  journal={IEEE Transactions on Neural Networks and Learning Systems},
  year={2025},
  publisher={IEEE}
}

@article{liu2025defense,
  title={Defense Strategy for Social Network Fake News Under Non-Real-Time Data Collection and Disjointed Decision Making},
  author={Liu, Zeyi and Zhang, Huaguang and Sun, Jiayue and Wang, Le},
  journal={IEEE Transactions on Network Science and Engineering},
  year={2025},
  publisher={IEEE}
}

@inproceedings{zhong2012comsoc,
  title={Comsoc: adaptive transfer of user behaviors over composite social network},
  author={Zhong, Erheng and Fan, Wei and Wang, Junwei and Xiao, Lei and Li, Yong},
  booktitle={Proceedings of the 18th ACM SIGKDD international conference on Knowledge discovery and data mining},
  pages={696--704},
  year={2012}
}

@inproceedings{rossi2015network,
  title={The network data repository with interactive graph analytics and visualization},
  author={Rossi, Ryan and Ahmed, Nesreen},
  booktitle={Proceedings of the AAAI conference on artificial intelligence},
  volume={29},
  number={1},
  year={2015}
}

@article{panzarasa2009patterns,
  title={Patterns and dynamics of users' behavior and interaction: Network analysis of an online community},
  author={Panzarasa, Pietro and Opsahl, Tore and Carley, Kathleen M},
  journal={Journal of the American Society for Information Science and Technology},
  volume={60},
  number={5},
  pages={911--932},
  year={2009},
  publisher={Wiley Online Library}
}

@article{barabasi1999emergence,
  title={Emergence of scaling in random networks},
  author={Barab{\'a}si, Albert-L{\'a}szl{\'o} and Albert, R{\'e}ka},
  journal={science},
  volume={286},
  number={5439},
  pages={509--512},
  year={1999},
  publisher={American Association for the Advancement of Science}
}

@article{cheng2025efficient,
  title={Efficient source detection in incomplete networks via sensor deployment and source approaching},
  author={Cheng, Le and Zhu, Peican and Tang, Keke and Gao, Chao and Wang, Zhen},
  journal={IEEE Transactions on Information Forensics and Security},
  year={2025},
  publisher={IEEE}
}

@article{bao2024graph,
  title={Graph contrastive learning for source localization in social networks},
  author={Bao, Qing and Jiang, Ying and Zhang, Wang and Jiao, Pengfei and Su, Jing},
  journal={Information Sciences},
  volume={679},
  pages={121090},
  year={2024},
  publisher={Elsevier}
}

@article{wu2020rank,
  title={Rank-one semidefinite programming solutions for mobile source localization in sensor networks},
  author={Wu, Xiaoping and Qi, Hengnian and Xiong, Naixue},
  journal={IEEE Transactions on Network Science and Engineering},
  volume={8},
  number={1},
  pages={638--650},
  year={2020},
  publisher={IEEE}
}

@article{pinto2012locating,
  title={Locating the source of diffusion in large-scale networks},
  author={Pinto, Pedro C and Thiran, Patrick and Vetterli, Martin},
  journal={Physical review letters},
  volume={109},
  number={6},
  pages={068702},
  year={2012},
  publisher={APS}
}

@inproceedings{chen2025structure,
  title={Structure-prior Informed Diffusion Model for Graph Source Localization with Limited Data},
  author={Chen, Hongyi and Ding, Jingtao and Liang, Xiaojun and Li, Yong and Zhang, Xiao-Ping},
  booktitle={Proceedings of the 34th ACM International Conference on Information and Knowledge Management},
  pages={250--259},
  year={2025}}

@inproceedings{huang2023two,
  title={Two-stage denoising diffusion model for source localization in graph inverse problems},
  author={Huang, Bosong and Yu, Weihao and Xie, Ruzhong and Xiao, Jing and Huang, Jin},
  booktitle={Joint European Conference on Machine Learning and Knowledge Discovery in Databases},
  pages={325--340},
  year={2023},
  organization={Springer}
}

@article{li2023rumor,
  title={Rumor source localization in social networks based on infection potential energy},
  author={Li, Weimin and Guo, Chang and Liu, Yanxia and Zhou, Xiaokang and Jin, Qun and Xin, Mingjun},
  journal={Information Sciences},
  volume={634},
  pages={172--188},
  year={2023},
  publisher={Elsevier}
}

@inproceedings{hou2025generalized,
  title={A generalized diffusion framework with learnable propagation dynamics for source localization},
  author={Hou, Dongpeng and Wang, Yuchen and Gao, Chao and Li, Xianghua},
  booktitle={Proceedings of the Thirty-Fourth International Joint Conference on Artificial Intelligence},
  pages={2919--2927},
  year={2025}
}

@article{ma2026rumor,
  title={Rumor source localization in social networks based on the propagation direction of observers},
  author={Ma, Zhi-Wei and Wang, Hong-Jue and Hu, Zhao-Long and Zhu, Xiang-Bin and Peng, Hao and L{\"u}, Lin-Yuan and Huang, Yi-Zhen and Li, Minglu},
  journal={Chaos: An Interdisciplinary Journal of Nonlinear Science},
  volume={36},
  number={1},
  year={2026},
  publisher={AIP Publishing}
}

@inproceedings{sun2025trace,
  title={Trace: Structural riemannian bridge matching for transferable source localization in information propagation},
  author={Sun, Li and Zhou, Suyang and Fang, Bowen and Zhang, Hechuan and Ye, Junda and Ye, Yutong and Yu, Philip S},
  booktitle={Proceedings of IJCAI},
  year={2025}
}

@article{zhou2025multi,
  title={Multi-Source Localization Based on Graph Representation Learning and Bayesian Optimization},
  author={Zhou, Zhangfei and Wang, Youguo and Zhai, Qiqing and Yan, Jun},
  journal={IEEE Transactions on Network Science and Engineering},
  volume={13},
  pages={4815--4832},
  year={2025},
  publisher={IEEE}
}

@article{ma2024dislpsi,
  title={DISLPSI: A framework for source localization in signed social networks with structural balance},
  author={Ma, Zhi-Wei and Wang, Hong-jue and Hu, Zhao-Long and Zhu, Xiang-Bin and Huang, Yi-Zhen and Huang, Faliang},
  journal={Physics Letters A},
  volume={523},
  pages={129772},
  year={2024},
  publisher={Elsevier}
}

@article{ma2024source,
  title={Source localization in signed networks with effective distance},
  author={Ma, Zhi-Wei and Sun, Lei and Ding, Zhi-Guo and Huang, Yi-Zhen and Hu, Zhao-Long},
  journal={Chinese Physics B},
  volume={33},
  number={2},
  pages={028902},
  year={2024},
  publisher={IOP Publishing}
}

@article{jiang2024source,
  title={Source localization in signed networks based on dynamic message passing algorithm},
  author={Jiang, Zhi-Xiang and Hu, Zhao-Long and Huang, Faliang},
  journal={Chaos, Solitons \& Fractals},
  volume={188},
  pages={115532},
  year={2024},
  publisher={Elsevier}
}

@article{cheng2022path,
  title={Path-based multi-sources localization in multiplex networks},
  author={Cheng, Le and Li, Xianghua and Han, Zhen and Luo, Tengyun and Ma, Lianbo and Zhu, Peican},
  journal={Chaos, Solitons \& Fractals},
  volume={159},
  pages={112139},
  year={2022},
  publisher={Elsevier}
}

@inproceedings{wang2024joint,
  title={Joint source localization in different platforms via implicit propagation characteristics of similar topics},
  author={Wang, Zhen and Hou, Dongpeng and Yin, Shu and Gao, Chao and Li, Xianghua},
  booktitle={Proceedings of the Thirty-Third International Joint Conference on Artificial Intelligence, IJCAI-24, Kate Larson (Ed.). International Joint Conferences on Artificial Intelligence Organization},
  pages={2424--2432},
  year={2024}
}

@article{gong2024hmsl,
  title={HMSL: Source localization based on higher-order Markov propagation},
  author={Gong, Chang and Li, Jichao and Qian, Liwei and Li, Siwei and Yang, Zhiwei and Yang, Kewei},
  journal={Chaos, Solitons \& Fractals},
  volume={182},
  pages={114765},
  year={2024},
  publisher={Elsevier}
}

@inproceedings{shi2025community,
  title={Community Partition-based Source Localization with Adaptive Observers Deployment},
  author={Shi, Jinchen and Fang, Yang and Tan, Zhen and Zhang, Xin and Zhao, Xiang},
  booktitle={Proceedings of the 34th ACM International Conference on Information and Knowledge Management},
  pages={2663--2673},
  year={2025}
}

@inproceedings{spinelli2017back,
  title={Back to the source: an online approach for sensor placement and source localization},
  author={Spinelli, Brunella and Celis, L Elisa and Thiran, Patrick},
  booktitle={Proceedings of the 26th international conference on world wide web},
  pages={1151--1160},
  year={2017}
}

@article{meng2025spreading,
  title={Spreading dynamics of information on online social networks},
  author={Meng, Fanhui and Xie, Jiarong and Sun, Jiachen and Xu, Cong and Zeng, Yutian and Wang, Xiangrong and Jia, Tao and Huang, Shuhong and Deng, Youjin and Hu, Yanqing},
  journal={Proceedings of the National Academy of Sciences},
  volume={122},
  number={4},
  pages={e2410227122},
  year={2025},
  publisher={National Academy of Sciences}
}

@article{fang2023comprehensive,
  title={A comprehensive survey on multi-view clustering},
  author={Fang, Uno and Li, Man and Li, Jianxin and Gao, Longxiang and Jia, Tao and Zhang, Yanchun},
  journal={IEEE Transactions on Knowledge and Data Engineering},
  volume={35},
  number={12},
  pages={12350--12368},
  year={2023},
  publisher={IEEE}
}

@inproceedings{liu2025coata,
  title={CoATA: Effective Co-Augmentation of Topology and Attribute for Graph Neural Networks},
  author={Liu, Tao and Lin, Longlong and Yu, Yunfeng and Ou, Xi and Zhang, Youan and Ye, Zhiqiu and Jia, Tao},
  booktitle={Proceedings of the 2025 International Conference on Multimedia Retrieval},
  pages={851--860},
  year={2025}
}

@article{he2025deciphering,
  title={Deciphering the overtaking phenomenon initiated by low-centrality nodes in spreading processes},
  author={He, Wenchao and Gao, Chao and Jia, Tao},
  journal={Communications Physics},
  volume={8},
  number={1},
  pages={266},
  year={2025},
  publisher={Nature Publishing Group UK London}
}

\begin{IEEEbiography}[{\includegraphics[width=1in,height=1.25in,clip,keepaspectratio]{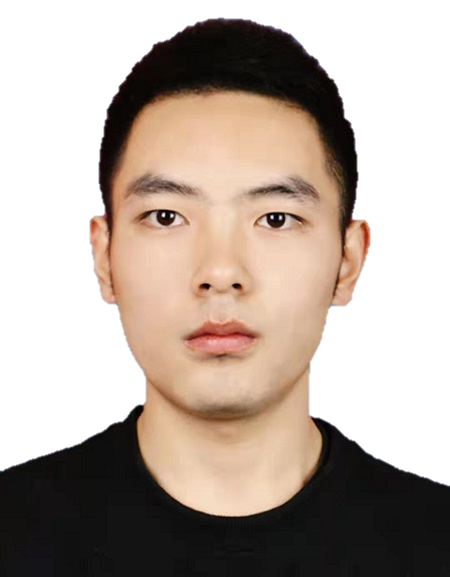}}]{Yansong Wang} received the MS degree from Southwest University, Chongqing, China. He is currently working toward the PhD degree with the College of Computer and Information Science, Southwest University, Chongqing, China. His research interests include social computing, information diffusion, and social recommendation.
\end{IEEEbiography}

\begin{IEEEbiography}[{\includegraphics[width=1in,height=1.25in,clip,keepaspectratio]{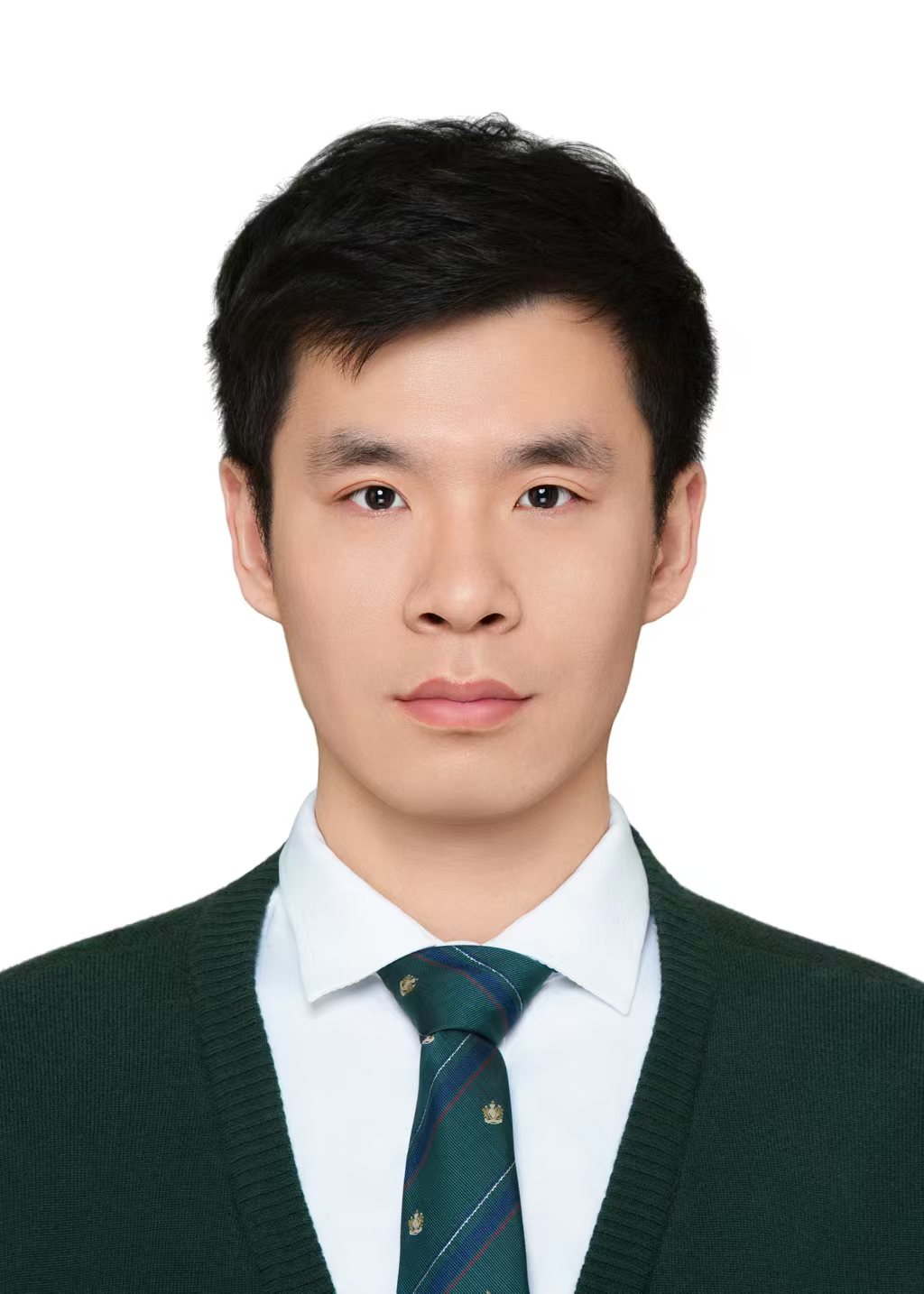}}]{Qisen Chai} received the MS degree from Southwest University, Chongqing, China, in 2023. He is currently working toward the PhD degree with the College of Computer and Information Science, Southwest University, Chongqing, China. His research interests include graph data mining and social computing.
\end{IEEEbiography}

\begin{IEEEbiography}[{\includegraphics[width=1in,height=1.25in,clip,keepaspectratio]{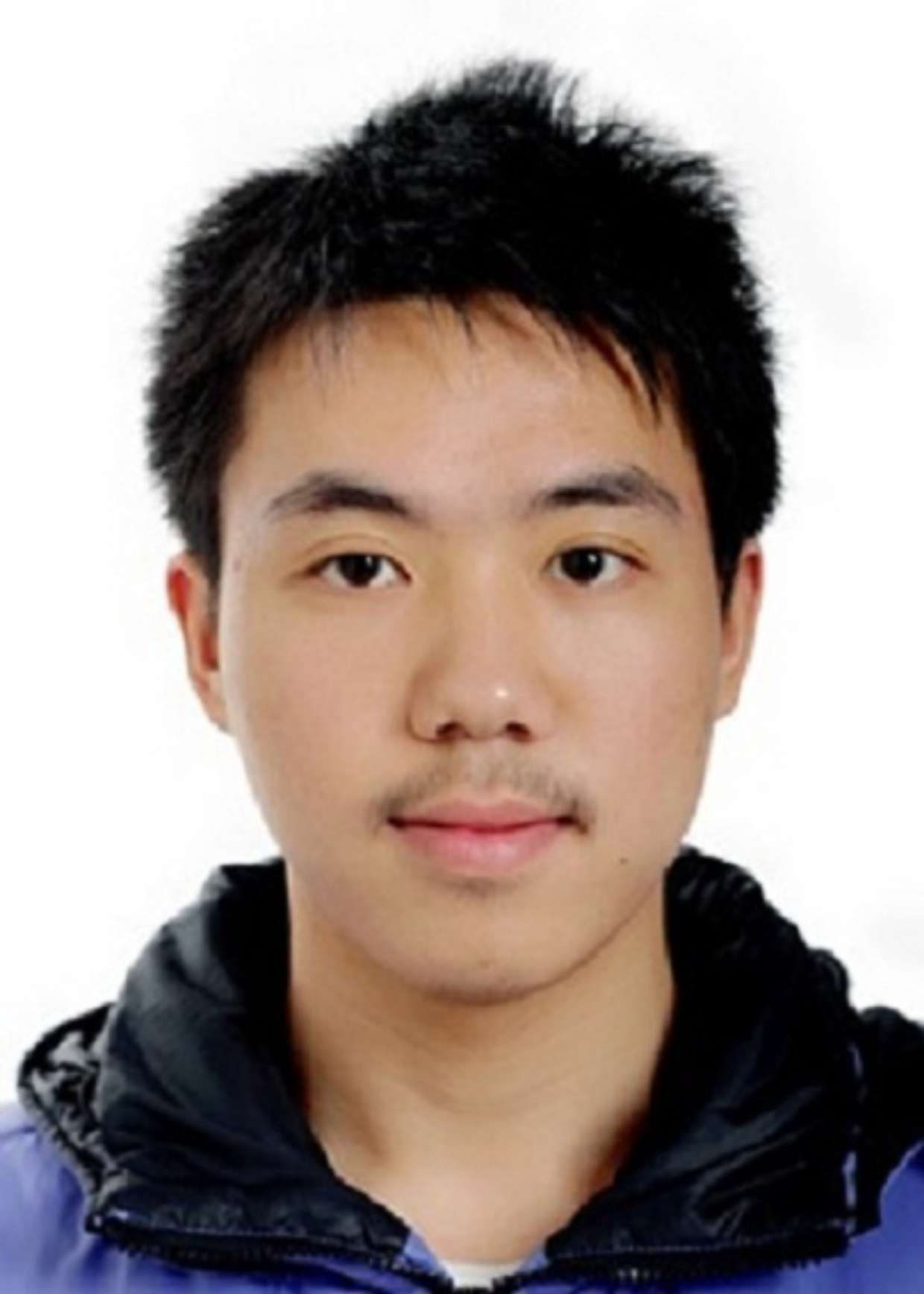}}]{Longlong Lin} received the PhD degree from Huazhong University of Science and Technology (HUST) in 2022. He is currently an associate professor in the College of Computer and Information Science, Southwest University, Chongqing, China. His current research interests include graph data management and mining, community search, graph clustering, and graph machine learning.
\end{IEEEbiography}

\begin{IEEEbiography}[{\includegraphics[width=1in,height=1.25in,clip,keepaspectratio]{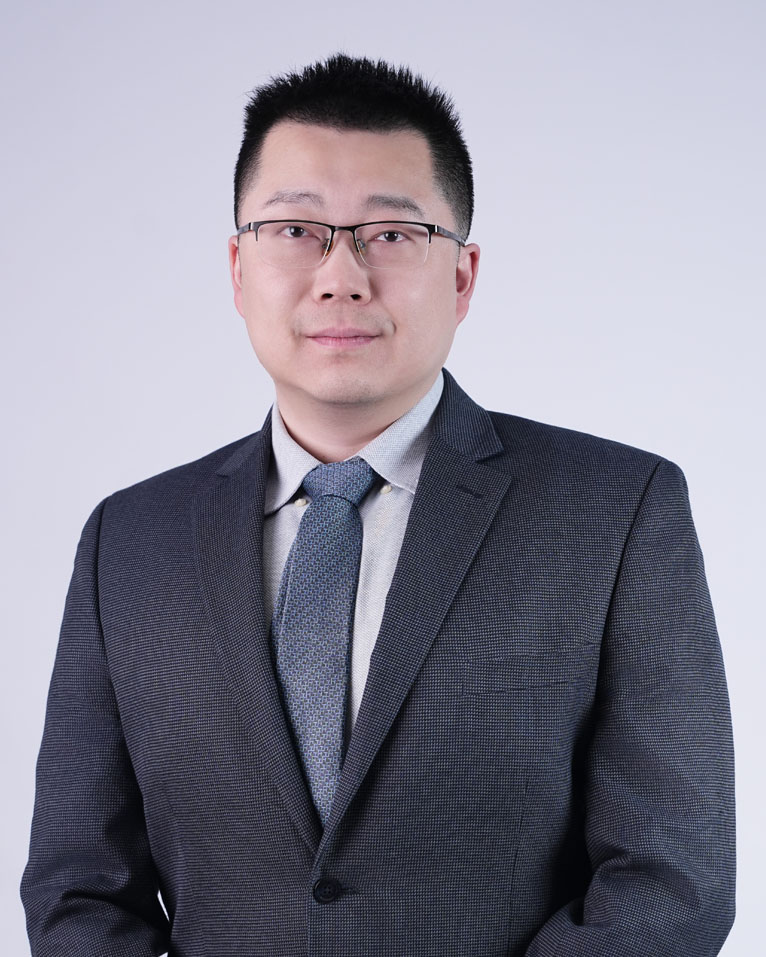}}]{Tao Jia} received the BSc degree from Nanjing University, China. He received his MSc and PhD degrees from Virginia Tech, USA. He is currently the Vice President of Chongqing Normal University, China. His research interests include graph mining, brain networks, and social computing.
\end{IEEEbiography}

\end{document}